\documentclass[12pt]{article}
\usepackage{epsfig}
\usepackage{amssymb}
\usepackage{amsmath}
\usepackage{amsfonts}

\newcommand{\R}{\mathbb{R}}
\newcommand{\C}{\mathbb{C}}

\newcommand{\fa}{\mathfrak{a}}
\newcommand{\fb}{\mathfrak{b}}

\newcommand{\fv}{\mathfrak{v}}

\def\RR{\mathbb{R}}

\def\CC{\mathbb{C}}

\newcommand{\cE}{\mathcal{E}}
\newcommand{\cH}{\mathcal{H}}
\newcommand{\cI}{\mathcal{I}}

\newcommand{\cO}{\mathcal{O}}
\newcommand{\cP}{\mathcal{P}}

\newcommand{\cT}{\mathcal{T}}
\newcommand{\cU}{\mathcal{U}}

\newcommand{\be}{\begin{equation}}
\newcommand{\ee}{\end{equation}}
\newcommand{\bea}{\begin{eqnarray}}
\newcommand{\eea}{\end{eqnarray}}
\newcommand{\nn}{\nonumber}
\newcommand{\kt}{\rangle}
\newcommand{\br}{\langle}

\newcommand{\ed}{\end{document}}

\newcommand{\rx}{{\rm x}}
\newcommand{\ry}{{\rm y}}
\newcommand{\rp}{{\rm p}}

\newcommand{\rH}{\mathbf{\mathrm{H}}}
\newcommand{\rh}{\mathbf{\mathrm{h}}}
\newcommand{\rX}{{\rm X}}
\newcommand{\rP}{{\rm P}}
\newcommand{\rE}{{\rm E}}

\newcommand{\bi}{\begin{itemize}}
\newcommand{\ei}{\end{itemize}}

\newcommand{\bce}{\begin{center}}
\newcommand{\ece}{\end{center}}

\begin{document}

\title{Application of Pseudo-Hermitian Quantum Mechanics to a
Complex Scattering Potential with Point Interactions }

\author{Hossein Mehri-Dehnavi\thanks{E-mail address:
mehri@alice.math.kindai.ac.jp}, Ali~Mostafazadeh\thanks{E-mail
address: amostafazadeh@ku.edu.tr}, and Ahmet Batal\thanks{Present
address: Department of Mathematics, Sabanc\i  University, 34956
Orhanli - Tuzla,  Istanbul, Turkey.}
\\
\\
$^*$~Research Center for Quantum Computing, Kinki University,
3-4-1 Kowakae,\\ Higashi-Osaka, Osaka 577-8502, Japan,\\
$^*$~Department of Physics, Institute for Advanced Studies in Basic
Sciences,\\ Zanjan 45195-1159, Iran,
\\
$^{\dagger~\ddagger}$~Department of Mathematics, Ko\c{c} University,
Rumelifeneri Yolu,\\ 34450 Sariyer, Istanbul, Turkey}
\date{ }
\maketitle

\begin{abstract}
We present a generalization of the perturbative construction of the
metric operator for non-Hermitian Hamiltonians with more than one
perturbation parameter. We use this method to study the
non-Hermitian scattering Hamiltonian:
$\rH=\rp^2/2m+\zeta_-\delta(\rx+\alpha)+\zeta_+\delta(\rx-\alpha)$,
where $\zeta_{\pm}$ and $\alpha$ are respectively complex and real
parameters and $\delta(\rx)$ is the Dirac delta function. For
regions in the space of coupling constants $\zeta_{\pm}$ where $\rH$
is quasi-Hermitian and there are no complex bound states or spectral
singularities, we construct a (positive-definite) metric operator
$\eta$ and the corresponding equivalent Hermitian Hamiltonian $\rh$.
$\eta$ turns out to be a (perturbatively) bounded operator for the
cases that the imaginary part of the coupling constants have
opposite sign, $\Im(\zeta_+)=-\Im(\zeta_-)$. This in particular
contains the $\cP\cT$-symmetric case: $\zeta_+=\zeta_-^*$. We also
calculate the energy expectation values for certain Gaussian wave
packets to study the nonlocal nature of $\rh$ or equivalently the
non-Hermitian nature of $\rH$. We show that these physical
quantities are not directly sensitive to the presence of
$\cP\cT$-symmetry.

 \end{abstract}
\vspace{5mm}

\vspace{2mm}


\section{Introduction}

A Hamiltonian operator $H$ is called $\cP\cT$-symmetric if it has
parity-time reversal symmetry, i.e., $[H,\cP\cT]=0$. Since the
publication of the pioneering work of Bender and Boettecher
\cite{bender-prl-98} non-Hermitian $\cP\cT$-symmetric Hamiltonians
have received much attention. This has led to the discovery of a
number of interesting theoretical
\cite{ruschhaupt,prl-09,gunther-prl-2008,ahmed,ahmet,bender-jmp-99}
as well as experimental \cite{pt-appl} implications of
$\cP\cT$-symmetric Hamiltonians. For extensive reviews see
\cite{bender-review,review} and references therein.

Perhaps the most prominent feature of a non-Hermitian
$\cP\cT$-symmetric Hamiltonian $H$ is that its spectrum is symmetric
about the real axis of the complex plane. In particular, if $H$ has
a discrete spectrum, either it is purely real or the nonreal
eigenvalues come in complex-conjugate pairs
\cite{bender-prl-98,bender-jmp-99}. It turns out that this is a
characteristic property of a wider class of non-Hermitian
Hamiltonian operators called the pseudo-Hermitian operators
\cite{p1,p2,p3}. A Hamiltonian $H$ is said to be pseudo-Hermitian if
its adjoint $H^\dag$ satisfies
  \bea
        H^\dag=\eta H \eta^{-1},
        \label{pseu}
  \eea
for some Hermitian invertible operator $\eta$.  Under the assumption
of the diagonalizability of $H$, one can show that its spectrum is
real if and only if there exists a positive-definite (metric)
operator $\eta$ satisfying the above equation \cite{review,p2,p3}.
In this case $H$ is called a quasi-Hermitian operator \cite{quasi}.

The diagonalizability of an operator is equivalent to the lack of
exceptional points and spectral singularities \cite{paper2}.
Exceptional points are degeneracy points where some of the
eigenvectors of the operator coalesce. This phenomenon have been the
subject of many theoretical \cite{gunther-prl-2008,EP-app2} and
experimental \cite{EP-app1} studies. It may appear for operators
acting in finite or infinite dimensional Hilbert spaces. In
contrast, spectral singularities can only appear for non-Hermitian
operators whose spectrum includes a real continuous part (See
\cite{guseinov} and references therein). Mathematically, they are
responsible for a break down of eigenfunction expansion
\cite{paper2}. Physically, they correspond to resonances having a
real energy (zero width) \cite{prl-09,ahmed,pra-09-2}.

As we mentioned above, a quasi-Hermitian Hamiltonian is a
diagonalizable operator with a completely real spectrum. This is not
however sufficient reason for using quasi-Hermitian operators as
observables in a quantum theory. This is because the
diagonalizability of an operator and the reality of its spectrum do
not necessarily imply the reality of the expectation values of the
operator. The latter condition is in fact equivalent to the
Hermiticity of the operator \cite{review}. The advantage of
quasi-Hermitian operators over other non-Hermitian operators is that
they can be made Hermitian by an appropriate modification of the
inner product on the Hilbert space. This is done using
positive-definite metric operators $\eta$ that satisfy (\ref{pseu}).
The modified inner product is given by $\br \cdot|\cdot \kt
_{\eta}:=\br \cdot|{\eta}\cdot \kt $, where $\br\cdot|\cdot\kt$ is
the inner product that defines the original Hilbert space $\cH$.
Endowing the vector space of state vectors with the inner product
$\br \cdot|\cdot \kt _{\eta}$, we define a new Hilbert space
$\cH_{{\rm phys}}$ in which $H$ acts as a Hermitian operator
\cite{review,jpa-2004b}. Hereafter we assume that $H$ is a
quasi-Hermitian operator and call $\cH_{{\rm phys}}$ the physical
Hilbert space.

In general, $\eta$ is not unique. This means that either one must
choose $\eta$ directly or fix it indirectly by demanding that a
so-called compatible irreducible set of quasi-Hermitian operators
will act as Hermitian operators in $\cH_{{\rm phys}}$, \cite{quasi}.
As explained in \cite{review}, the latter approach is very difficult
to implement in practice, if the only available information is the
form of the quasi-Hermitian operator $H$. This is because to select
the members of a compatible irreducible set of quasi-Hermitian
operators containing $H$, we need to construct the most general
metric operator $\eta$ fulfilling (\ref{pseu}). For a
quasi-Hermitian Schr\"odinger operator $H=-\frac{d^2}{dx^2}+v(x)$
with a typical complex potential, this is an extremely difficult
open problem. Dealing with this problem is particularly difficult
for complex scattering potentials such as the one studied in the
present article, because the continuous spectrum of $H$ is doubly
degenerate. In what follows, we will assume that a choice for $\eta$
and consequently $\cH_{{\rm phys}}$ is made a priori.

Because $\eta$ is positive-definite, it has a unique
positive-definite square root $\rho:=\sqrt{\eta}:\cH\rightarrow
\cH$. It is easy to show that $\rho:\cH_{\rm phys}\rightarrow \cH$
is a unitary operator, i.e.,
    \be
    \br \rho \cdot | \rho \cdot \kt =\br \cdot|\cdot \kt _{\eta}.
    \label{rho-unitary}
    \ee
It establishes the unitary equivalence of the (Hilbert space,
Hamiltonian) pairs: $(\cH_{\rm phys},H)$ and $(\cH,h)$ where
$h:=\rho H\rho^{-1}$, \cite{jpa-2004b,jpa-2003}. The operator $h$ is
a Hermitian operator acting in the original Hilbert space $\cH$. It
is called the equivalent Hermitian Hamiltonian associated with the
metric operator $\eta$, \cite{jpa-2008b}. $(\cH_{\rm phys},H)$ and
$(\cH,h)$  provide equivalent representations of the same quantum
system. They are respectively called the pseudo-Hermitian and
Hermitian representations \cite{review,jpa-2004b}.

In the  pseudo-Hermitian representation we work with the physical
Hilbert space $(\cH_{\rm phys},\br \cdot|\cdot\kt_\eta)$, and the
quasi-Hermitian observables $H, X:=\rho^{-1}x\rho,
P:=\rho^{-1}p\rho, \cdots$, where $x, p, \cdots$ are the usual
Hermitian observables. In the Hermitian representation we work with
the usual Hilbert space $\cH$ and the Hermitian observables $h:=\rho
H\rho^{-1}, x, p, \cdots$. A particle that is described by the state
vector $|\psi\kt\in\cH_{\rm phys}$ and the Hamiltonian $H$ can also
described by the state vector $\rho|\psi\kt\in\cH$ and the
Hamiltonian $h$, \cite{review,jpa-2004b}.

Working with the Hermitian representation has the advantage of
revealing the underlying classical system. This is of great
importance to derive the physical meaning of the system
$\cite{jpa-2005,jpa-2006a}$ and establish a classical-to-quantum
correspondence principle. In order to employ the Hermitian
representation, we need to compute the equivalent Hermitian
Hamiltonian $h$. This in turn requires the calculation of $\rho$. A
well-known method of constructing $h$ is to use the exponential
representation $\eta=e^{-Q}$ for the metric operator and apply the
perturbation scheme developed in \cite{jpa-2005,bender-prd-2004}. We
will begin our analysis by extending this method for the cases that
the Hamiltonian $H$ involves more than one perturbation parameter.
We will then apply this method to treat the quantum system defined
by the double-delta function potential:
    \be
     v(x)=z_-\delta(x+a)+z_+\delta(x-a),~~~~z_\pm\in\C,~a\in\R^+.
     \label{ddelta-fn}
     \ee
The spectral properties of this and analogous complex point
interaction potentials have been considered in \cite{DD-potential}.
See also \cite{jones-prd-2008,jones-znojil}. A thorough
investigation of (\ref{ddelta-fn}) that addresses the issue of the
presence of spectral singularities and provided means for locating
the regions in the space $M$ of coupling constants where the
Hamiltonian is quasi-Hermitian is conducted in \cite{paper2}.

In the present paper we offer an explicit construction of an
appropriate metric operator for $H$ in a three-dimensional subspace
of $M$ that includes the ${\cP\cT}$-symmetric potentials of the type
(\ref{ddelta-fn}) with $z_-=z_+^*$. Using this metric operator we
determine the equivalent Hermitian Hamiltonians and compute energy
expectation values for some Gaussian wave packets. Our results are
valid irrespective of the presence of ${\cP\cT}$-symmetry.
Therefore, they allow for a direct examination of the physical
consequences of imposing ${\cP\cT}$-symmetry.

\section{Spectral Representation of Metric Operator for
a Scattering Potential}

Consider a non-Hermitian Hamiltonian $H$ with a complex-valued
scattering potential $v(x):\RR\rightarrow \CC$. Suppose that $v(x)$
depends linearly on a set $\vec{z}:=(z_1,z_2,\cdots ,z_d)$ of
complex coupling constants so that $H^{\dag}$ can be obtained by
replacing $\vec{z}$ with $\vec{z}^*$ in $v(x)$. A typical example is
the double-delta function potential (\ref{ddelta-fn}). For the cases
that $H$ has no bound states (square-integrable eigenfunctions), its
spectrum is doubly degenerate and its eigenvalue equation takes the
form
    \be
    H|\psi_{\fa,k}^{\vec z}\kt=k^2|\psi_{\fa,k}^{\vec z}\kt.
    \label{eig-value}
    \ee
Here $\fa\in\{1,2\}$ is the degeneracy label and $k\in\RR^+$ is the
spectral label. In the absence of spectral singularities $H$ is a
diagonalizable operator, and we can use $|\psi_{\fa,k}^{\vec z}\kt$
together with a set of (generalized) eigenvectors
$|\phi_{\fa,k}^{\vec z}\kt$ of $H^{\dag}$ to construct a complete
biorthonormal system, i.e.,
    \be
    \br\phi_{\fa,k}^{\vec z}|\psi_{\fb,q}^{\vec
    z}\kt=\delta_{\fa\fb}\delta(k-q),~~~~\sum_{\fa=1}^{2}
    \int_{0}^{\infty}|\psi_{\fa,q}^{\vec z}\kt\br\phi_{\fa,k}^{\vec
    z}|dk=1.
    \label{biortho-set}
    \ee
In this case the following formula gives a (positive-definite)
metric operator for the Hamiltonian $H$ \cite{cqg-2003}.
    \be
    \eta=\sum_{\fa=1}^{2}\int_{0}^{\infty}|\phi_{\fa,
    k}^{\vec{z}}\kt\br \phi_{\fa, k}^{\vec{z}}|\:dk.
    \label{metric}
    \ee
It is easy to see that $\eta$ satisfies (\ref{pseu}).\footnote{Here
we assume that $H$ has no bound sates. If there are $N$ bound sates
with real energies, (\ref{metric}) is generalized to
$\eta=\sum_{\fa=1}^{{\fa}_j}\sum_{j=1}^{N}|\phi_{\fa,
    j}^{\vec{z}}\kt\br \phi_{\fa, j}^{\vec{z}}|+\sum_{\fa=1}^{2}\int_{0}^{\infty}|\phi_{\fa,
    k}^{\vec{z}}\kt\br \phi_{\fa, k}^{\vec{z}}|dk,$
where ${\fa}_j$ denotes to the degree of degeneracy of the $j$-th
eigenvalue.} We can use this operator to define the
positive-definite inner product
$\br\cdot|\cdot\kt_{\eta}:=\br\cdot|{\eta}\cdot\kt$ and the
corresponding Hilbert space $\cH_{\rm phys}$ in which $H$ acts as a
Hermitian operator.

Next, we recall that because of the arbitrariness in the choice of
the biorthonormal system (in particular
$|\phi_{\fa,k}^{\vec{z}}\kt$), the metric operator (\ref{metric}) is
not unique \cite{jmp-2003}. In what follows we will try to choose
the biorthonormal system
$\{|\psi_{\fa,k}^{\vec{z}}\kt,|\phi_{\fa,k}^{\vec{z}}\kt\}$ in such
a way that in the Hermitian limit, where all the coupling constants
are real, the metric operator~(\ref{metric}) tends to unity.

Since $H^{\dag}$ can be found by replacing $\vec{z}$ with
$\vec{z}^*$ in $H$, the simplest way to find a set of eigenvectors
of $H^\dagger$ is to replace $\vec{z}$ with $\vec{z}^*$ in
$|\psi_{\fa,k}^{\vec z}\kt$. It is however not difficult to see that
in general $\{|\psi_{\fa,k}^{\vec
z}\kt,|\psi_{\fa,k}^{\vec{z}^*}\kt\}$ does not satisfy
(\ref{biortho-set}). In fact, one can show that
    \be
    \left(\begin{array}{cc}
    \br\psi^{\vec z^*}_{1,k}|\psi^{\vec z}_{1,q}\kt &
    \br\psi^{\vec z^*}_{1,k}|\psi^{\vec z}_{2,q}\kt\\
    \br\psi^{\vec z^*}_{2,k}|\psi^{\vec z}_{1,q}\kt &
    \br\psi^{\vec z^*}_{2,k}|\psi^{\vec
    z}_{2,q}\kt\end{array}\right)=\delta(k-q)K
    \label{phi-psi-2}
    \ee
where $K=(K_{\fa \fb})$ is a $2\times 2$ matrix that is generally
different from the identity matrix.

One way of constructing a metric operator with appropriate Hermitian
limit (namely identity operator) is to find a matrix
$\cU(\vec{z};k)$ satisfying
    \be
    \cU^{\dag}(\vec{z}^*;k)K(\vec{z};k)\cU(\vec{z};k)=I_{2\times
    2}
    \label{Uab}
    \ee
and use the biorthonormal system $\{|\tilde\psi_{\fa,k}^{\vec
z}\kt,|\tilde\phi_{\fa,k}^{\vec{z}}\kt\}$ defined by
    \be
    |\tilde{\phi}_{\fa, k}^{\vec{z}}\kt:=
    \cU_{\fa \fb}(\vec{z}^*;k)|\psi_{\fb,
    k}^{\vec{z^*}}\kt,~~~~
    |\tilde{\psi}_{\fa, k}^{\vec{z}}\kt:=
    \cU_{\fa \fb}(\vec{z};k)|\psi_{\fb,
    k}^{\vec{z}}\kt.
    \label{symmetric-eigen}
    \ee
This approach relies on the solution of Eq.~(\ref{Uab}). It is clear
form (\ref{phi-psi-2}) that
    \be
    K^{\dag}(\vec{z}^*;k)=K(\vec{z};k).
    \ee
In view of this identity we can rewrite (\ref{Uab}) as
     \be
     \cU^{\dag}(\vec{z}^*;k)\:{K^{\dag}}(\vec{z}^*;k)^{\frac{1}{2}}\:
     {K}(\vec{z};k)^{\frac{1}{2}}\:\cU(\vec{z};k)=I_{2\times 2}.
    \label{Uab1}
    \ee
This equation has infinitely many solutions. Perhaps the simplest
solution is
    \be
    \cU={K}^{-\frac{1}{2}}.
    \label{U-solution}
    \ee
In general, $K^{-\frac{1}{2}}$ has an extremely complicated form.
This leads to serious computation difficulties in the perturbative
expansion of the metric operator. Furthermore, there is no assurance
that this choice of the biorthonormal system yields a bounded metric
operator.

In Ref.~\cite{jpa-2006b} this method is employed to calculate a
metric operator for a delta function potential,
$v(x)=z\,\delta(\rx)$, with a complex coupling constant $z$ having a
positive real part. In this case, it yields a perturbatively bounded
metric operator. We will discuss the possibility of applying this
method for the complex double-delta function potential
(\ref{ddelta-fn}) in Section~4.

\section{The Perturbative Expansion of $h$ }

Let $H:\cH\to\cH$ be a quasi-Hermitian Hamiltonian of the form
    \bea
        H=H^{(0)}+H^{(1)},~~~~H^{(1)}:=\sum_{i=1}^{d}z_i\,H_i
        \label{perturb h0}
    \eea
where $H^{(0)},H_1,\cdots,H_d$ are Hermitian operators and
$z_1,z_2,\cdots ,z_d\in\C$ are complex parameters. Suppose that
$|z_i|\ll 1$, for all $i\in\{1,2,\cdots,d\}$, so that we can use
them as perturbation parameters. Then, (\ref{perturb h0}) is a
perturbative expansion of $H$, with $H^{(0)}$ and $H^{(1)}$
respectively denoting the zeroth and the first order terms.

Consider the perturbative expansion of an arbitrary operator $A$
depending on $z_1,z_2,\cdots ,z_d$. Let $n_1,n_2,\cdots n_d$ be
non-negative integers and $N:=n_1+n_2+\cdots+n_d$. Then we call the
sum of terms proportional to $z_1^{n_1}z_2^{n_2}\cdots z_d^{n_d}$
``the $N$-th order term'' of this expansion and denote it by
$A^{(N)}$. We also use $\cO(z^N)$ to label the sum of the terms of
order greater than or equal to $N$.

Because the first order term of the Hamiltonian is generally
non-Hermitian, we write it as
    \bea
        H^{(1)}=H^{(1)}_{h.}+H^{(1)}_{a.h.},
        \label{perturb h1}
    \eea
where  $H^{(1)}_{h.}$ and $H^{(1)}_{a.h.}$ stand for the Hermitian
and anti-Hermitian parts of $H^{(1)}$, respectively. In view of
quasi-Hermiticity of $H$, there is a positive-definite metric
operator $\eta$ satisfying $H^\dag=\eta H \eta^{-1}$. This relation
together with Eqs.~(\ref{perturb h0}) and (\ref{perturb h1}) imply
    \bea
        \eta H \eta^{-1}=H^{(0)}+H^{(1)}_{h.}-H^{(1)}_{a.h.}.
        \label{pseudo}
    \eea
Our aim is to use this equation to construct a perturbative
expansion for a metric operator with correct Hermitian limit:
    \bea
         \eta={ 1}+\eta^{(1)}+\eta^{(2)}+\cO(z^3),
        \label{eta expan}
    \eea
and the corresponding equivalent Hermitian Hamiltonian $h$.

First, we recall that because $\eta$ is a positive-definite
operator, there is a Hermitian operator $Q$ satisfying
\cite{bender-prd-2004}
    \bea
    \eta=e^{-Q}.
    \label{eta}
    \eea
Next, we use (\ref{pseudo}), (\ref{eta}), the
Baker-Campbell-Hausdorff identity, and the perturbative expansion of
$Q$, namely
    \bea
    Q=Q^{(1)}+Q^{(2)}+\cO(z^3),
    \label{q expansion}
    \eea
to obtain \cite{jpa-2004b}:
    \bea
        H^\dag&=&H+[H,Q]+\frac{1}{2}[[H,Q],Q]+\cO(z^3)\label{H dag2}\\
        &=&H^{(0)}+H^{(1)}+[H^{(0)},Q^{(1)}]+[H^{(0)},Q^{(2)}]+[H^{(1)},Q^{(1)}]+
        \frac{1}{2}[[H^{(0)},Q^{(1)}],Q^{(1)}]+\cO(z^3).\nn
            \eea
Comparing the right hand side of (\ref{pseudo}) with the first order
term of the last equation, we have
    \bea
        H^{(1)}_{h.}-H^{(1)}_{a.h.}=H^{(1)}_{h.}+H^{(1)}_{a.h.}+[H^{(0)},Q^{(1)}],
        \label{H first}
    \eea
or equivalently
    \bea
        [H^{(0)},Q^{(1)}]=-2H_{a.h.}^{(1)}.
        \label{H relation first}
    \eea
It is also clear  from (\ref{pseudo}) that the second order term of
$H^\dag$ vanishes. This implies that the second order term in
(\ref{H dag2}) must satisfy
    \bea
        [H^{(1)},Q^{(1)}]+[H^{(0)},Q^{(2)}]+
        \frac{1}{2}[[H^{(0)},Q^{(1)}],Q^{(1)}]=0.
        \label{H second}
    \eea

Similarly, using
    \be
    \rho=e^{-Q/2}
    \label{rho}
    \ee
we find the perturbative expansion of the equivalent Hermitian
Hamiltonian $h$:
    \bea
        h&=&\rho H
        \rho^{-1}=H+[H,\mbox{$\frac{Q}{2}$}]+
        \frac{1}{2}[[H,\mbox{$\frac{Q}{2}$}],
        \mbox{$\frac{Q}{2}$}]+\cO(z^3)
        \label{h-her}\\
        &=&H^{(0)}+H^{(1)}+\frac{1}{2}\left\{[H^{(0)},Q^{(1)}]+[H^{(1)},Q^{(1)}]+[H^{(0)},Q^{(2)}]
        +\frac{1}{4}[[H^{(0)},Q^{(1)}],Q^{(1)}]\right\}+\cO(z^3).
        \nn
        \eea
In light of (\ref{H relation first}) and (\ref{H second}), we can
simplify this expression as follows.
    \bea
        h&=&H^{(0)}+H^{(1)}_{h.}-
        \frac{1}{8}\, [[H^{(0)},Q^{(1)}],Q^{(1)}]
        +\cO(z^3)=
        H^{(0)}+H^{(1)}_{h.}+\frac{1}{4}[H^{(1)}_{a.h.},Q^{(1)}]+\cO(z^3).
        \label{h1}
       \eea
This is a straightforward generalization of the results of
Refs.~\cite{jpa-2005} to the cases involving more than one complex
perturbation parameter. See also \cite{FF-2006}.

Next, we use the identity $Q^{(1)}=-\eta^{(1)}$ to express the
integral kernel $h({ x},{y}):=\br {x}|h| {y}\kt $ of the equivalent
Hermitian Hamiltonian. This yields {\small
    \bea
   h({ x},{ y})&=&
         H^{(0)}({ x},{ y})+H^{(1)}_{h.}({ x},{ y})-
        \frac{1}{4}\br { x}|[H^{(1)}_{a.h.},\eta^{(1)}]| { y}\kt
          +\cO(z^3)\label{green h}\\
         &=& H^{(0)}({ x},{ y})+H^{(1)}_{h.}({ x},{ y})
         -\frac{1}{4}\int_{\R}\left(H^{(1)}_{a.h.}({ x},{ y}')\eta^{(1)}({ y}',{ x})
        -\eta^{(1)}({ x},{ y}')H^{(1)}_{a.h.}({ y}',{ x})\right) d { y}'
          +\cO(z^3).\nn
               \eea}

Having obtained the equivalent Hermitian Hamiltonian, we can examine
the physical content of the model using its Hermitian
representation. Alternatively, we can purse the study of this system
in its pseudo-Hermitian representation. This requires the
construction of the pseudo-Hermitian observables $O=\rho^{-1}o\rho$,
where $o=x,p,\cdots$ are the usual Hermitian observables. Following
a similar approach to the one leading to (\ref{h-her}), we find
    \bea
    O&=&\rho^{-1}o\rho=
    o-[o,\mbox{$\frac{Q}{2}$}]+\frac{1}{2}[[o,\mbox{$\frac{Q}{2}$}],
    \mbox{$\frac{Q}{2}$}]+\cO(z^3)\nn\\
    &=& o -\frac{1}{2}\left\{[o,Q^{(1)}]+[o,Q^{(2)}]-
    \frac{1}{4}[[o,Q^{(1)}],Q^{(1)}]
    \right\}+\cO(z^3).
    \label{pseu-observable}
    \eea

\section{The Double-Delta Function Potential}
\subsection{Eigenfunctions and the $K$-matrix}

In this subsection we summarize some of the properties of the
double-delta function potential that are reported in
Ref.~\cite{paper2}.

Consider the time-independent Schr\"odinger equation:
    \be
    \rH \,\psi=\left[-\frac{\hbar^2}{2m}\frac{d^2}{d\rx^2}+
    \zeta_+\delta(\rx-\alpha)+\zeta_-\delta(\rx+\alpha)\right]\psi=
    \rE\,\psi.
    \label{eg-va}
    \ee
Let $\ell$ be an arbitrary length scale. Defining the dimensionless
quantities
    \bea
    z_\pm:=\frac{2m\ell\zeta_\pm}{\hbar^2},~~~x:=\frac{\rx}{\ell},~~~
    a:=\frac{\alpha}{\ell},
    ~~~E:=\frac{2m\ell^2\rE}{\hbar^2}
    \label{dimentionless parameters}
    \eea
we express the corresponding dimensionless Hamiltonian as
        \bea
        H:=\frac{2m\ell^2\rH}{\hbar^2}=
    -\frac{d^2}{dx^2}+z_+\delta(x-a)+z_-\delta(x+a).
    \label{H}
    \eea
We can easily solve the eigenvalue problem for $H$ and obtain the
following expression for the eigenvectors $\psi_{1,k}^{\vec z}$ of
the Hamiltonian $H$.
    \bea
    \psi_{1,k}^{\vec z}(x)&=&\frac{1}{\sqrt{2\pi}}
    \Big\{e^{ikx}-\frac{i z_-}{2k}\left[
    e^{-ik(x+2a)}-e^{ikx}\right]\theta(-x-a)
    -\frac{i z_+}{2k}\left[
    e^{ikx}-e^{-ik(x-2a)}\right]\theta(x-a)\Big\},~~~~
    \label{psi1}\\
    \psi_{2,k}^{\vec z}(x)&=&\psi_{1,-k}^{\vec z}(x),~~~~
    \label{psi2}
    \eea
where $\theta(x):=[{\rm sign}(x)+1]/2$ is the step function.
Clearly,
    \bea
    \psi_{1,k}^{\vec z^*}(x)&=&\frac{1}{\sqrt{2\pi}}
    \Big\{e^{ikx}-\frac{i z^*_-}{2k}\left[
    e^{-ik(x+2a)}-e^{ikx}\right]\theta(-x-a)
    -\frac{i z^*_+}{2k}\left[
    e^{ikx}-e^{-ik(x-2a)}\right]\theta(x-a)\Big\},
    \label{phi1}~~~~\\
    \psi_{2,k}^{\vec z^*}(x)&=&\psi_{1,-k}^{\vec z^*}(x),~~~~
    \label{phi2}
    \eea
and the entries of the matrix $K$ of Eq.~(\ref{phi-psi-2}) takes the
form \cite{paper2}:
    \bea
    K_{11}(k)&=&K_{22}(k)=1+\frac{z_-^2+z_+^2}{4k^2},
    \label{e-K11}\\
    K_{12}(k)&=&(4k^2)^{-1}\left[
    iz_-(2k-iz_-)e^{2iak}-iz_+(2k+iz_+)e^{-2iak}\right],
    \label{e-K12}\\
    K_{21}(k)&=& K_{12}(-k).
    \label{e-K21}
    \eea
As we can infer from these equations, the matrix $K^{-1/2}$ has a
rather complicated expression. This makes a direct application of
the method of Section 2 extremely difficult. In what follows we will
try to pursue a different approach for constructing a metric
operator with a correct Hermitian limit. We will also demand that,
at least to the first order of perturbation, the metric operator be
densely-defined and bounded \cite{review}.

Similarly to the case of the single delta function potential,
$v(x)=z\,\delta(\rx)$, the fact that we have exact and closed-form
expressions for the eigenfunctions of the Hamiltonian (\ref{H}) does
mean that we can perform an exact calculation of a metric operator.
Note that the latter requires choosing an appropriate set of
eigenfunctions $\phi^{\vec z}_{\fa k}$ and evaluating the integral
in (\ref{metric}) exactly. Lack of a systematic method of selecting
$\phi^{\vec z}_{\fa k}$ (alternatively the matrix-valued function
${\cal U}$ appearing in (\ref{Uab})) is the main reason why we
conduct a perturbative analysis of the problem.

\subsection{Perturbative Calculation of the Metric Operator}

For the cases that $\Re(z_{\pm})>0$ and $|\epsilon_{\pm}|
:=\left|\frac{\Im(z_{\pm})}{\Re(z_{\pm})}\right|\ll 1$, the
Hamiltonian $H$ is a quasi-Hermitian operator \cite{paper2}.
Therefore $\epsilon_{\pm}:=\frac{\Im(z_{\pm})}{\Re(z_{\pm})}$ may be
employed as perturbation parameters. As shown in Appendix A, this
choice leads to a metric operator that does not tend to the identity
operator in the Hermitian limit ($\epsilon_{\pm}\to 0$). An
alternative choice for perturbation parameters is the coupling
constants $(z_+,z_-)=:\vec{z}$ themselves. In the remainder of this
section we construct a metric operator $\eta$ using these
perturbation parameters.

According to (\ref{e-K11}) - (\ref{e-K21}), the zeroth order term of
the matrix $K$ obtained for the eigenvectors (\ref{psi1}) -
(\ref{phi2}) is the identity matrix. This in turn implies that the
zeroth order term for the corresponding metric operator is the
identity operator. Yet this metric operator is plagued with the
problem of unboundedness and the lack of a dense domain.

In order to construct a densely-defined and bounded metric operator,
we use the following ansatz for the eigenvectors of $H^\dagger$.
    \be
    |{\phi}_{\fa, k}^{\vec{z}}\kt:=
    |\psi_{\fa, k}^{\vec{z^*}}\kt +
    z_{-}^*\sum_{\fb=1}^{2}
     u_{-,\fb}(k)
     |{\psi}_{\fb, k}^{\vec{z^*}}\kt+
     z_{+}^*\sum_{\fb=1}^{2}u_{+,\fb}(k)
     |{\psi}_{\fb, k}^{\vec{z^*}}\kt,
    \label{phi-till-general}
    \ee
where  $u_{\pm,\fb}(k)$'s are free weight functions. In Appendix B
we describe a procedure for selecting a proper set of weight
functions. We could do this successfully only for the special case
that the coupling constants differ by a sign: $z_+=-z_-=:z$. In this
case, we find (see Appendix B)
 \bea
    {\phi}_{1, k}^{\vec{z}}(x)&=&\left(1-\frac{z^*}{2ik}\right){\psi}_{1,
    k}^{\vec{z^*}}(x)+\frac{z^* \cos 2ak }{2ik} {\psi}_{2,
    k}^{\vec{z^*}}(x)+\cO(z^2),
    \label{phi-prim-1}\\
    {\phi}_{2, k}^{\vec{z}}(x)&=&{\phi'}_{1,-k}^{\vec{z}}(x),
    \label{phi-prim-2}\\
    {\eta}(x,y)&=&\delta(x-y)+{\eta}^{(1)}(x,y)+\cO(z^2),
    \label{bounded-metric-1}\\
    {\eta}^{(1)}(x,y)&=&
    \frac{i\Im(z)}{2}{\rm
    sign}(x-y)[\theta(x^++y^+)-\theta(x^{-}+y^{-})]
    \label{bounded-metric}
    \eea
where
    \bea
    x^{\pm}:=x\pm a,~~~~y^{\pm}:=y\pm a.
    \label{x-y-pm}
    \eea
The metric operator (\ref{bounded-metric-1}) has the following
desirable properties.
    \begin{enumerate}
    \item It tends to the identity operator in the Hermitian limit.
    \item It is a densely-defined bounded operator.
    \item It satisfies the differential equation for the (pseudo-) metric
    operators associated with pseudo-Hermitian Schr\"odinger operators
    \cite{jmp-2006}.
    \item For the $\cP\cT$-symmetric case corresponding to a purely
    imaginary $z$, it reduces to the metric operator obtained
    in \cite{ahmet}.
    \end{enumerate}

\begin{figure}[t]
    \begin{center}
   \includegraphics[scale=.64,clip]{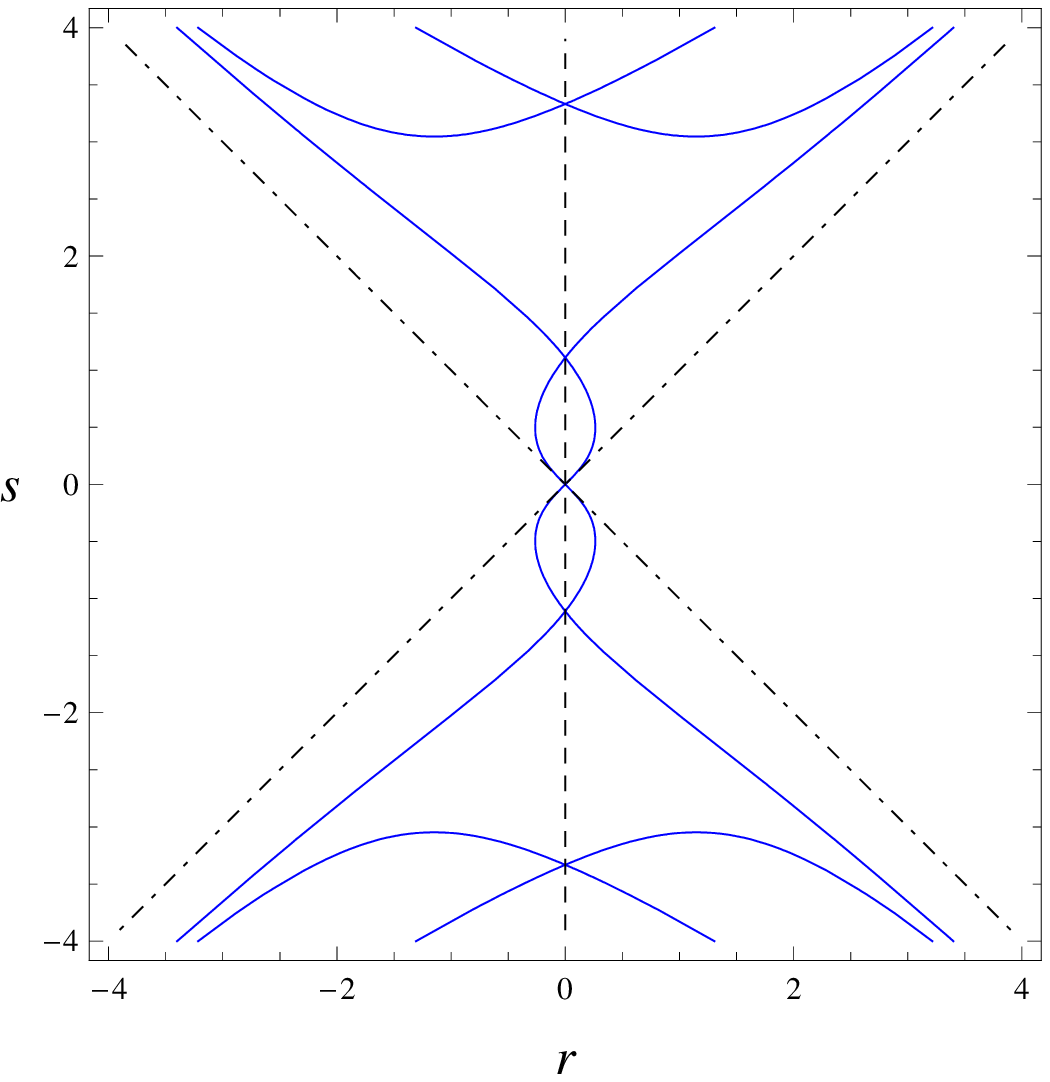}\hspace{1cm}
    \includegraphics[scale=.12,clip]{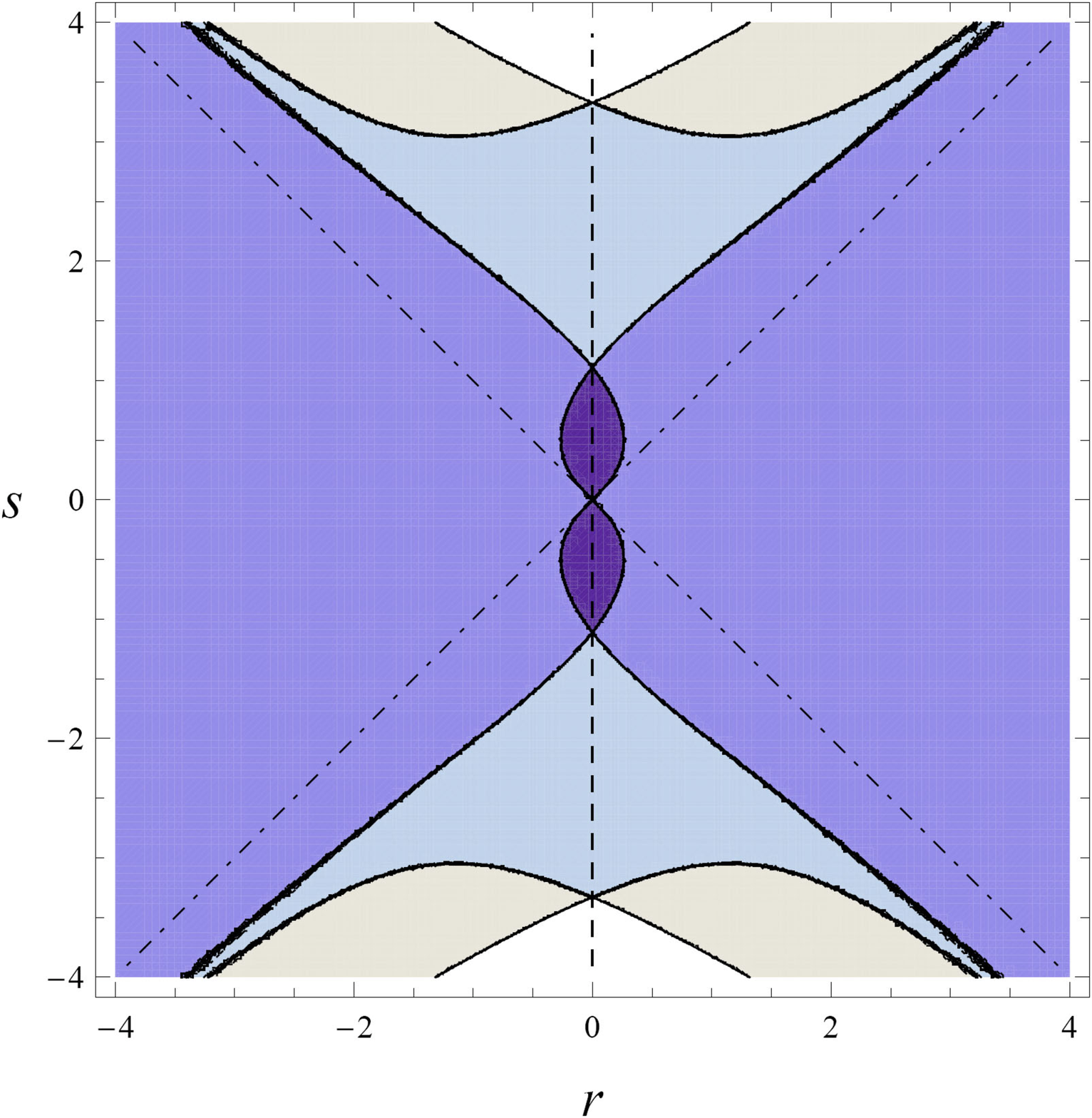}
    \parbox{15cm}{\caption{The figure on left shows the curves in
    the $az$-plane along which $H$ has spectral
    singularities. Here $r:=a\Re(z)$ and $s:=a\Im(z)$. The vertical
    dashed lines correspond to the $\cP\cT$-symmetric case with purely
    imaginary $z$. The diagonal lines are the lines $s=\pm r$. The figure on
    the right shows the number of bound states (generally complex
    eigenvalues with square-integrable eigenfunctions). As the
    color changes from the darkest to lightest the number of bound sates
    will changes from zero to four.
    \label{fig1}}}\end{center}
    \end{figure}

We would like to emphasize that the above construction is valid only
for the cases that the Hamiltonian is quasi-Hermitian. Otherwise the
metric operator will not satisfy the pseudo-Hermiticity relation
(\ref{pseu}). Therefore, it is of outmost importance to determine
the range of valued of $z$ for which the Hamiltonian is
quasi-Hermitian. These are the regions where $H$ has no spectral
singularities or complex eigenvalues. Figure~\ref{fig1} shows the
regions in the complex $az$-plane where the Hamiltonian has spectral
singularities and bound states. This figure is obtained using the
contour integral method described in Ref.~\cite{paper2}.
For small values of $|z|$ with $|\Re(z)|>|\Im(z)|$, $H$ has a bound
state (an eigenvalue with a square-integrable eigenfunction). This
corresponds to a real eigenvalue if and only if $|\Im (z)|=0$,
\cite{paper2}. In the darkest region shown in the right-hand figure
the Hamiltonian is free of spectral singularities and bound states.
This is a region where it is quasi-Hermitian, and
(\ref{bounded-metric-1}) provides a reliable metric operator.

\subsection{Equivalent Hermitian Hamiltonian}

Inserting (\ref{bounded-metric}) in (\ref{green h}) and doing the
necessary calculations, we obtain the following expression for the
equivalent Hermitian operator $h$ defined by the metric operator
(\ref{bounded-metric-1}).
    \bea
     h(x,y)& = & \delta(x-y)\left(-\frac{d^2}{dx^2}+\Re(z)\left[\delta(x^-)
       -\delta(x^+)\right]\right) \label{h-bounded}\\
       &&\hspace{-2.4cm}\begin{array}{c}
         +\mbox{\large$\frac{\Im(z)^2}{8}$}\Big{\{}
      \delta(x^+)[\theta(y^+)-\theta(y-3a)]+ \delta(x^-)[\theta(y+3a)-\theta(y^-)]\\
         \hspace{3.4cm}+\delta(y^+)[\theta(x^+)-\theta(x-3a)] + \delta(y^-)[
     \theta(x+3a)-\theta(x^-)]\Big{\}}+\cO(z^3).
       \end{array}
       \nn
    \eea
If we multiply $h(x,y)$ by $\frac{\hbar^2}{2m \ell^3}$ and use
(\ref{dimentionless parameters}), we obtain the dimensionful
Hermitian Hamiltonian operator:
      \bea
        \rh(\rx,\ry)& =&\delta(\rx-\ry)\left(- \frac{\hbar^2}{2m }\frac{d^2}{d\rx^2}+\Re(\zeta)\left[\delta(\rx-\alpha)
       -\delta(\rx+\alpha)\right]\right) \label{h-bounded-dim} \\
     &&\hspace{-2.6cm}\begin{array}{cc}
      & +\mbox{\large$\frac{m\Im(\zeta)^2}{4\hbar^2}$}\Big{\{}
      \delta(\rx+\alpha)[\theta(\ry+\alpha)-\theta(\ry-3\alpha)]+ \delta(\rx-\alpha)[\theta(\ry+3\alpha)-\theta(\ry-\alpha)] \\
       &\hspace{3.6cm}+\delta(\ry+\alpha)[\theta(\rx+\alpha)-\theta(\rx-3\alpha)] + \delta(\ry-\alpha)[
     \theta(\rx+3\alpha)-\theta(\rx-\alpha)]\Big{\}}+\cO(\zeta^3),
     \end{array}
           \nn
    \eea
where $\zeta=\zeta_+=\frac{\hbar^2 z}{2 m \ell}$. The second order
(nonlocal) part of the Hamiltonian (\ref{h-bounded-dim}) is
equivalent to the anti-Hermitian part of the Hamiltonian (\ref{H}).
In the following we study the effect of this nonlocal part on the
energy expectation value, $\rE_{\psi}:=\br\psi|\rh|\psi\kt$, for a
particle described by a normalized Gaussian position wave function
$\psi\in L^2(\R)$.

The action of $\rh$ on an arbitrary element $\psi$ of the Hilbert
space $\cH$ is given by
    \bea
     \br \rx|\rh|\psi\kt &=& - \frac{\hbar^2}{2m }\psi''(\rx)+\Re(\zeta)\left[\delta(\rx-\alpha)\psi(\alpha)
      -\delta(\rx+\alpha)\psi(-\alpha)\right] \label{h-on-psi}\\
      &&+\frac{m[\Im(\zeta)]^2}{4\hbar^2}\Big{\{}[\theta(\rx+\alpha)-\theta(\rx-3\alpha)]\psi(-\alpha)
      +[\theta(\rx+3\alpha)-\theta(\rx-\alpha)]\psi(\alpha)
      \nn\\
      &&\hspace{2cm} +\delta(\rx+\alpha)\int_{-\alpha}^{3\alpha}\psi(\ry)d\ry+ \delta(\rx-\alpha)
      \int_{-3\alpha}^{\alpha}\psi(\ry)d\ry\Big{\}}+\cO(\zeta^3)\nn.
      \eea
The  first line of this Equation coincides with the action of the
Hermitian part of the original quasi-Hermitian Hamiltonian, namely
$\rH_{h.}=\rH^{(0)}+\rH^{(1)}_{h.}$.

In view of (\ref{h-on-psi}),
    \bea
    \rE_{\psi}:=\br\psi|\rh|\psi\kt&= &
    \frac{\hbar^2}{2m}\int_{-\infty}^{\infty}|\psi'(\rx)|^2d\rx+\Re(\zeta)
    \left[|\psi(\alpha)|^2-|\psi(-\alpha)|^2\right]
    \label{E-general}\\
    &&+\frac{m[\Im(\zeta)]^2}{2\hbar^2}~\Re\left(\psi^*(-\alpha)\int_{-\alpha}^{3\alpha}\psi(\rx)d\rx
    +\psi^*(\alpha)\int_{-3\alpha}^{\alpha}\psi(\rx)d\rx\right)+
    \cO(\zeta^3),
    \nn
    \eea
where $\psi$ is a normalized wave function. A typical example is a
Gaussian wave packet,
     \bea
    \psi_1(\rx)=\frac{1}{(\pi \sigma^2)^{1/4}}\exp\left(
    -\frac{\rx^2}{2\sigma^2}+ik\rx\right),
      \label{pssi1}
       \eea
with mean position $\br \rx\kt_{\psi_1}:=\br\psi_1|\rx|\psi_1\kt=0$
and mean momentum $\br
\rp\kt_{\psi_1}:=\br\psi_1|\rp|\psi_1\kt=\hbar k$. Substituting
(\ref{pssi1}) in (\ref{E-general}), we find
    \bea
    \rE_{\psi_1}=
    \frac{\hbar^2}{4m}\left(2k^2-\sigma^{-2}\right)+
    \frac{m[\Im(\zeta)]^2}{\sqrt{2}\hbar^2}\:
    U(\alpha,\sigma,k)+\cO(\zeta^3),
    \label{E1}
    \eea
where
    \be
    U(\alpha,\sigma,k):=\exp\left(-\frac{\alpha^2+k^2\sigma^4}{2
    \sigma^2}\right)\Re \left\{ e^{-ik\alpha}
    \left[{\rm erf}
    \left(\frac{ik\sigma^2+3\alpha}{\sqrt{2}\sigma}\right)-
    {\rm erf}\left(\frac{ik\sigma^2-\alpha}{\sqrt{2}\sigma}\right)
    \right] \right\}
    \label{U}
    \ee
describes the effect of the nonlocal part of $\rh$ (equivalently the
non-Hermitian part of $\rH$), and ${\rm
erf}(\rx):=\frac{2}{\sqrt{\pi}}\int_0^\rx e^{-t^2}dt$ is the error
function.
    \begin{figure}[ht]\begin{center}
    \includegraphics[scale=.12,clip]{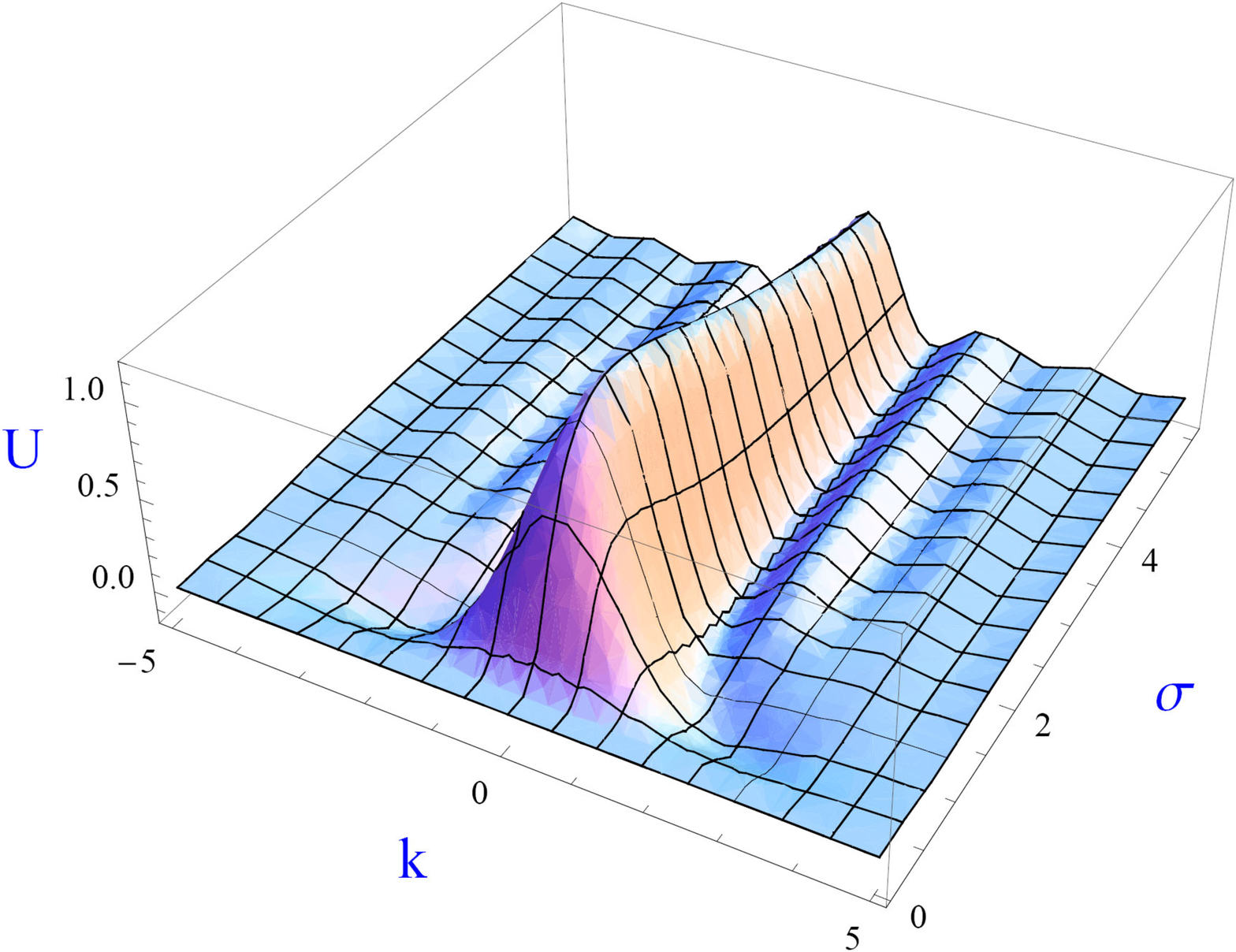}\hspace{1cm}
    \includegraphics[scale=.11,clip]{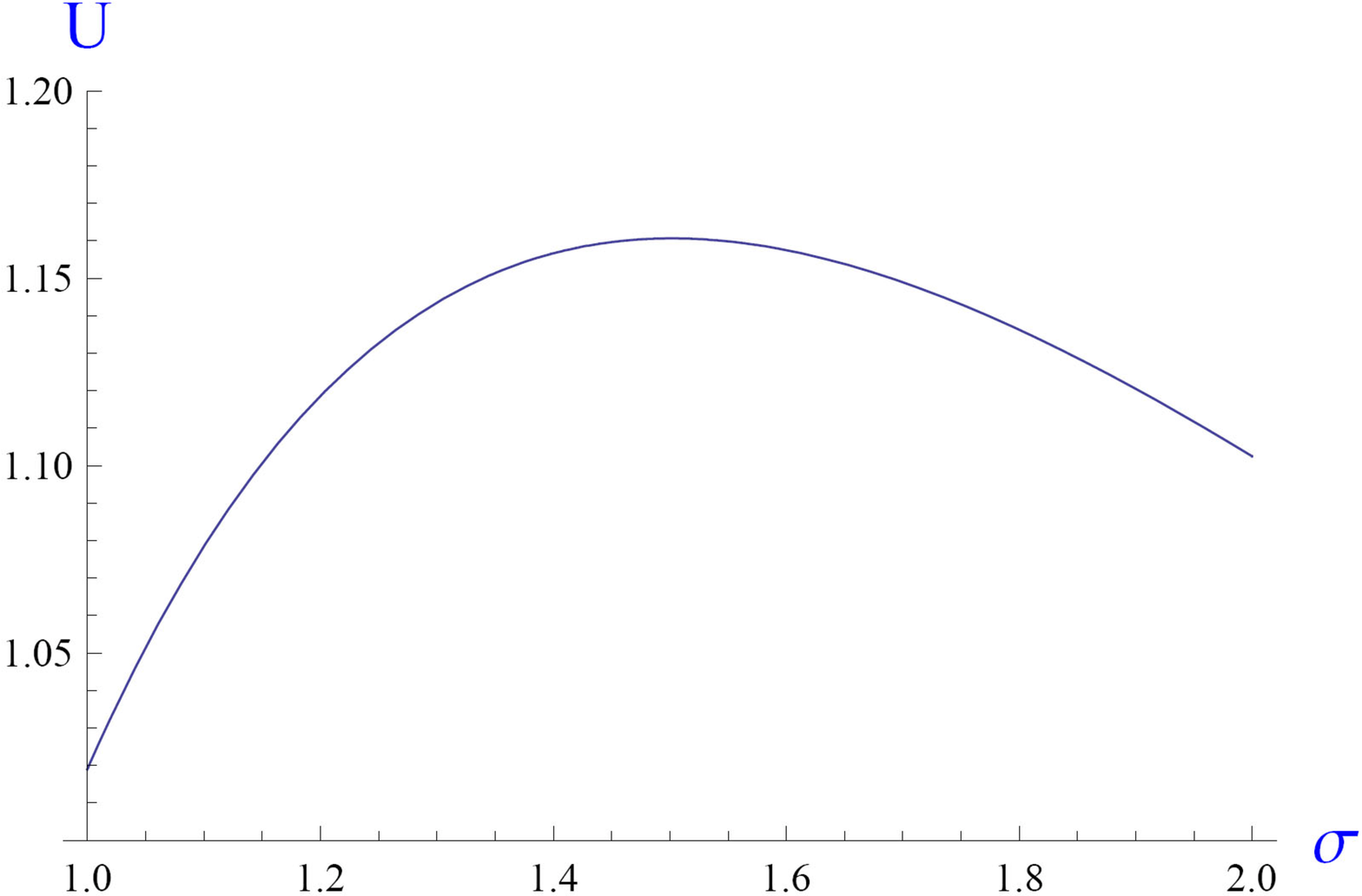}
    \parbox{15cm}{\caption{Plots of $U(1,\sigma,k)$ (on the left) and
    $U(1,\sigma,0)$ (on the right).
    \label{fig2}}}\end{center}
\end{figure}
Figure \ref{fig2} shows the plots of $U(\alpha,\sigma,k)$ and
$U(\alpha,\sigma,0)$ for $\alpha=1$. It turns out that the
non-Hermiticity effect attains its maximum around
$(\sigma,k)=(1.5\alpha,0)$ and decays rapidly for the mean momentums
$\br \rp\kt_{\psi_1}=\hbar k$ outside  the range
$\left(-\frac{\hbar}{\alpha},\frac{\hbar}{\alpha}\right)$.

Next, we compute the energy expectation value for a stationary
Gaussian wave packet with mean position $\br\rx\kt_{\psi_2}=\rx_0$:
   \bea
    \psi_2(\rx)
    =\frac{1}{(\pi \sigma^2)^{1/4}}\exp\left(
    -\frac{(\rx-\rx_0)^2}{2\sigma^2}\right).
      \label{pssi2}
       \eea
In view of (\ref{E-general}), we have
     \bea
     E_{\psi_2}&=&\frac{\hbar^2}{4m\sigma^{2}}+ \frac{\Re[\zeta]}{\sigma
    \sqrt{\pi}}V(\alpha,\sigma,\rx_0)+\frac{m[\Im(\zeta)]^2}{2\sqrt{2}\hbar^2}W(\alpha,\sigma,\rx_0)+\cO(\zeta^3),
     \label{E2}\\
    V(\alpha,\sigma,\rx_0)&:=&\exp\left(
    -\sigma^{-2}(\rx_0-\alpha)^2\right)-\exp\left(
    -\sigma^{-2}(\rx_0+\alpha)^2\right)\\
    W(\alpha,\sigma,\rx_0)&:=&\exp \left({-\frac{(\alpha+\rx_0)^2}{2
    \sigma^2}}\right)\left\{{\rm
    erf}\left(\frac{\alpha+\rx_0}{\sqrt{2}\sigma}\right)+
        {\rm erf}\left(\frac{3\alpha-\rx_0}{\sqrt{2}\sigma}\right) \right.\label{W}\\
    &&\hspace{3.7 cm} +\left.e^{\frac{2\alpha\rx_0}{\sigma^2}}
    \left[{\rm erf}\left(\frac{\alpha-\rx_0}{\sqrt{2}\sigma}\right)+
    {\rm erf}\left(\frac{3\alpha+\rx_0}{\sqrt{2}\sigma}\right)
    \right]\right\}.\nn
         \eea
Here $W(\alpha,\sigma,\rx_0)$ reflects the effect of the nonlocal
part of $\rh$ (non-Hermitian part of $\rH$).

\begin{figure}[ht]\begin{center}
    \includegraphics[scale=.15,clip]{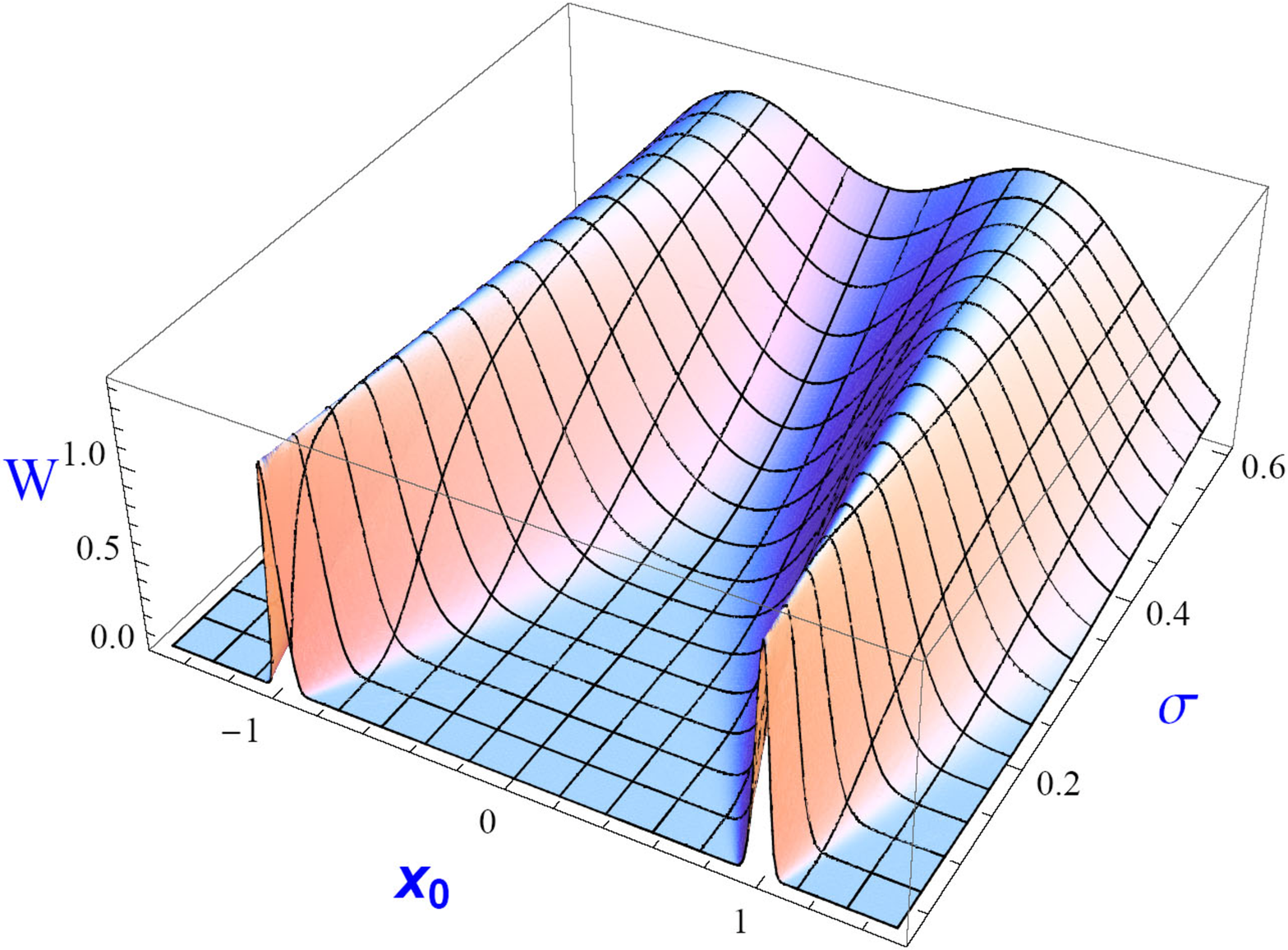}\hspace{1cm}
    \includegraphics[scale=.12,clip]{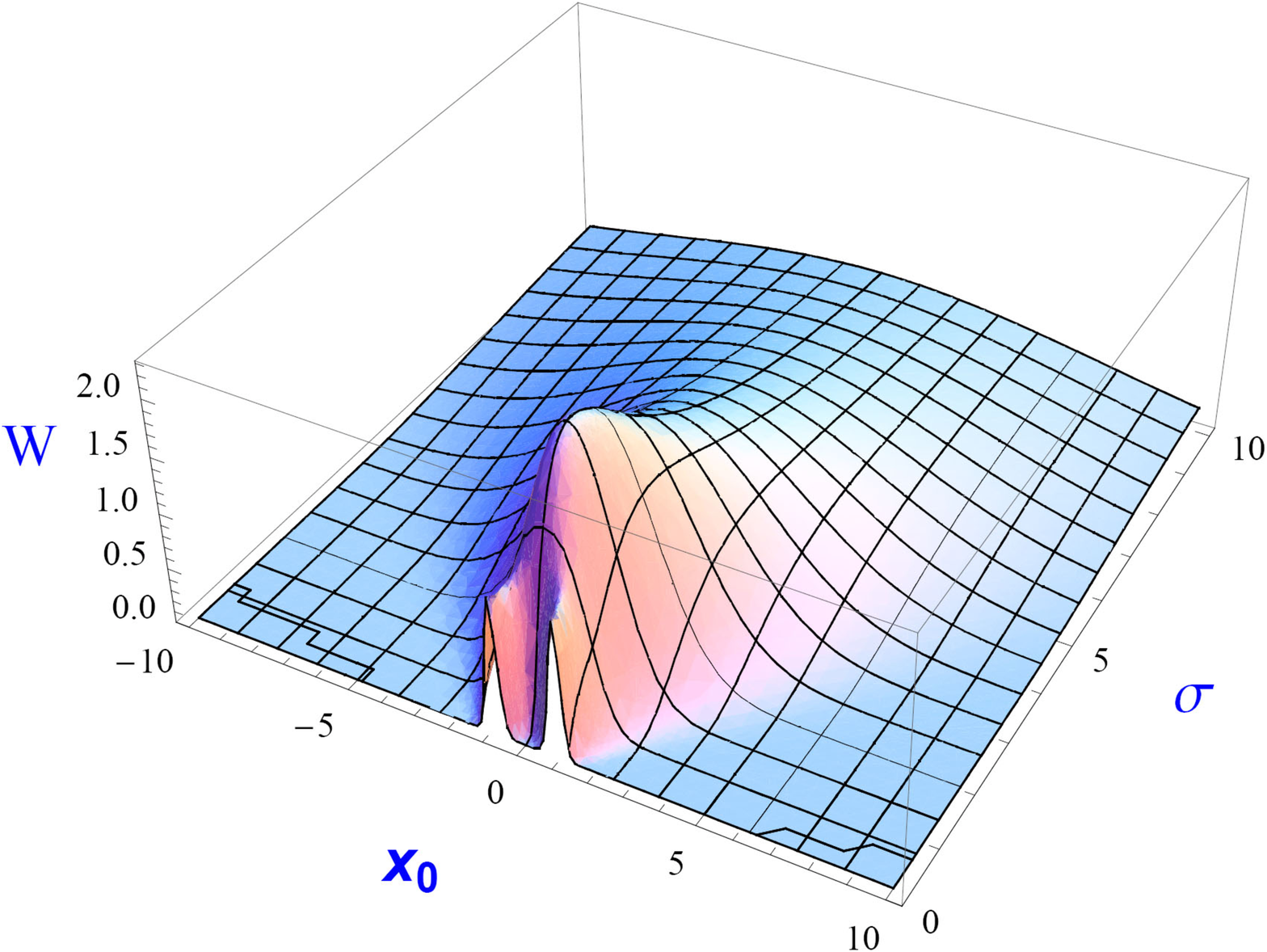}
    \parbox{15cm}{\caption{Plots of $W(\alpha,\sigma,\rx_0)$  for $\sigma\in[0.03,0.6]$
    (on the left) and $\sigma\in[0.2,10]$ (on the right) in units where
    $\alpha=1$.
    \label{fig3}}}\end{center}
    \end{figure}

Figure \ref{fig3} shows the plots of $W(1,\sigma,\rx_0)$ for
$\sigma\in[0.03,0.6]$ and $\sigma\in[0.2,10]$. The non-Hermitian
effect attains its maximum at $(\rx_0\approx 0,\sigma\approx 1.5
\alpha)$. For small values of $\sigma$, it persists for wave packets
with mean position belonging to open intervals
$(\pm\alpha-1.5\sigma,\pm\alpha+1.5\sigma)$. Outside these intervals
it decays rapidly.

\subsection{Pseudo-Hermitian Position and Momentum Operators}

To calculate the dimensionless pseudo-Hermitian position and
momentum operators corresponding to the metric operator
(\ref{bounded-metric-1})-(\ref{bounded-metric}), we substitute $x$
and $p$ for $o$ in (\ref{pseu-observable}) and use
$Q^{(1)}=-\eta^{(1)}$ and the identities:
    $$\br x|[x,A]|y\kt=(x-y)\br x|A|y \kt,~~~~
    \br x|[p,A]|y\kt=-i(\partial_x-\partial_y)\br x|A|y \kt,$$
where $A$ is a linear operator. This yields
    \bea
    \br x|X|y\kt&=&\br x|x|y\kt
    +\frac{i\Im(z)}{4}|x-y|
    \left[\theta(x^++y^+)-\theta(x^-+y^-)\right]+\cO(z^2),\\
    \br x|P|y\kt&=&\br x|p|y\kt
    +\Im(z)\delta(x-y)\left[\theta(x^+)-\theta(x^-)\right]+\cO(z^2).
    \eea

Next we obtain the dimensionful pseudo-Hermitian position
($\rX:=\ell X$) and momentum ($\rP:=\frac{\hbar}{\ell}P$) operators:
    \bea
    \br \rx|\rX|\ry\kt&=&\br \rx|\rx|\ry\kt
    +\frac{im\Im(\zeta)}{2\hbar^2}|\rx-\ry|
    \left[\theta(\rx+\ry+2\alpha)-
    \theta(\rx+y-2\alpha)\right]+\cO(\zeta^2),\\
    \br \rx|\rP|\ry\kt&=&\br \rx|\rp|\ry\kt
    +\frac{2m\Im(\zeta)}{\hbar}\delta(\rx-\ry)
    \left[\theta(\rx+\alpha)-\theta(\rx-\alpha)\right]+\cO(\zeta^2).
    \eea

\subsection{Calculating Metric for More General Cases}

In Section~4.2, we constructed a metric operator with the desired
Hermitian limit for the cases that the coupling constants
$\zeta_\pm$ differed by a sign. Our construction was based on the
spectral method that yielded $\eta$ in terms of its spectral
decomposition. An alternative method of constructing a metric
operator for a Schr\"odinger operator,
$-\frac{\hbar^2}{2m}\,\frac{d^2}{d\rx^2}+v(\rx)$, is the one based
on the universal differential equation \cite{jmp-2006}:
    \be
    \left(-\partial_{\rx}^2+\partial_{\ry}^2+\frac{2m}{\hbar^2}\left[v^{*}(\rx)-v(\ry)\right]\right)\eta(\rx,\ry)=0.
    \label{metric-DE}
    \ee
In this section we use this method to extend the results of the
preceding sections to a more general class of double-delta function
potentials.

Consider a quasi-Hermitian Hamiltonian operator,
    \be
    \rH_1=\rH^{(0)}+\rH^{(1)}_{a.h.}=-\frac{\rp^2}{2m}+v_1(\rx),
    \label{H1-def}
    \ee
with a purely imaginary potential $v_1=i\Im\left(v_1\right)$, and a
corresponding metric operator $\eta_1$ satisfying (\ref{metric-DE})
with $v=v_1$. Let $\zeta$ be a complex perturbation parameter such
that $v_1$ is proportional to $\Im(\zeta)$. This suggests the
following perturbative expansion for $\eta$.
    \be
    \eta_1=1+\eta^{(1)}_1+\cO(\zeta^2).
    \ee

Next, suppose that the potential $v_1$ is supplemented with a real
part $v_2$ that is proportional to $\Re(\zeta)$. Then it is easy to
see that up to the first order of perturbation, $\eta_1$ satisfies
(\ref{metric-DE}) for the potential $v_1(\rx)+v_2(\rx)$, i.e.,
$\eta_1$ is a metric operator also for the Hamiltonian
    \be
    \rH_2=\rH_1+\rH^{(1)}_{h.}=-\frac{\rp^2}{2m}+v_1(\rx)+v_2(\rx).
    \label{H2-def}
    \ee
Furthermore, in light of (\ref{h1}), the equivalent Hermitian
Hamiltonian for $\rH_1$ and $\rH_2$ are respectively given by
    \bea
    \rh_1&=&\rH^{(0)}-\frac{1}{4}[\rH^{(1)}_{a.h.},\eta^{(1)}_{1}]+\cO(\zeta^3)\nn\\
    \rh_2&=&\rH^{(0)}+\rH^{(1)}_{h.}-\frac{1}{4}[\rH^{(1)}_{a.h.},\eta^{(1)}_{1}]+\cO(\zeta^3)=\rh_1+\rH^{(1)}_{h.}+\cO(\zeta^3).
    \label{h1-h2}
    \eea

Now, we confine our attention to the double-delta function
potential. In Section~4.2, we constructed an appropriate metric
operator, namely (\ref{bounded-metric-1}), for a double-delta
function potential whose couplings differed by a sign. In view of
the argument given in the preceding paragraph, it is also a valid
metric operator for the more general case that the real part of the
coupling constants are arbitrary (but small) and their imaginary
part differ by a sign:
    \be
    \rH=-\frac{\rp^2}{2m}+\zeta_+\delta(\rx-\alpha)+\zeta_-\delta(\rx+\alpha),~~\Im(\zeta_+)=-\Im(\zeta_-).
    \label{general-h}
    \ee
Another class of potentials that admit (\ref{bounded-metric-1}) as
an appropriate metric operator is \cite{jones-prd-2008}:
    \be
    v(\rx)=-\xi\delta(\rx)+i\lambda
    \left[\delta(\rx-\alpha)-\delta(\rx+\alpha)\right],~~~~~\xi\in
    \RR^+,~~~\lambda\in\RR.
    \ee

Next, we use the metric operator (\ref{bounded-metric-1}) to compute
the equivalent Hermitian Hamiltonian $\rh$ for the Hamiltonian
(\ref{general-h}). Using (\ref{h1-h2}) and performing the necessary
calculations, we find
    \bea
       \rh(\rx,\ry)& =&\delta(\rx-\ry)\left(- \frac{\hbar^2}{2m }\frac{d^2}{d\rx^2}+\Re(\zeta_+)\delta(\rx-\alpha)
      +\Re(\zeta_-)\delta(\rx+\alpha)\right)\label{h-bounded-dim-g2} \\
      &&\hspace{-2.6cm}\begin{array}{cc}
      & +\frac{m[\Im(\zeta_+)]^2}{4\hbar^2}\Big{\{}
      \delta(\rx+\alpha)[\theta(\ry+\alpha)-\theta(\ry-3\alpha)]+ \delta(\rx-\alpha)[\theta(\ry+3\alpha)-\theta(\ry-\alpha)] \\
       &\hspace{3.6cm}+\delta(\ry+\alpha)[\theta(\rx+\alpha)-\theta(\rx-3\alpha)] + \delta(\ry-\alpha)[
      \theta(\rx+3\alpha)-\theta(\rx-\alpha)]\Big{\}}+\cO(\zeta^3).
      \end{array}\nn
    \eea
As we expected the nonlocal part of $\rh$ is identical with the one
obtained for the special case where coupling constants differ by a
sign.

The general $\cP\cT$-symmetric double-delta function potential
($\zeta_+=\zeta_-^*=:\zeta$) corresponds to a special case of the
Hamiltonian~(\ref{general-h}). Our analysis shows that, within the
framework of perturbation theory, the physical effects associated
with the non-Hermiticity of the Hamiltonian $\rH$, that are
contained in the nonlocal part of the equivalent Hermitian
Hamiltonian $\rh$, are not sensitive to the presence of
$\cP\cT$-symmetry. This is because by perturbing the Hermitian part
of the Hamiltonian we may destroy its $\cP\cT$-symmetry while
preserving the same non-Hermitian (nonlocal) effects on the
physically observables quantities like energy expectation values.

\begin{figure}[t]
    \begin{center}
   \includegraphics[scale=.64,clip]{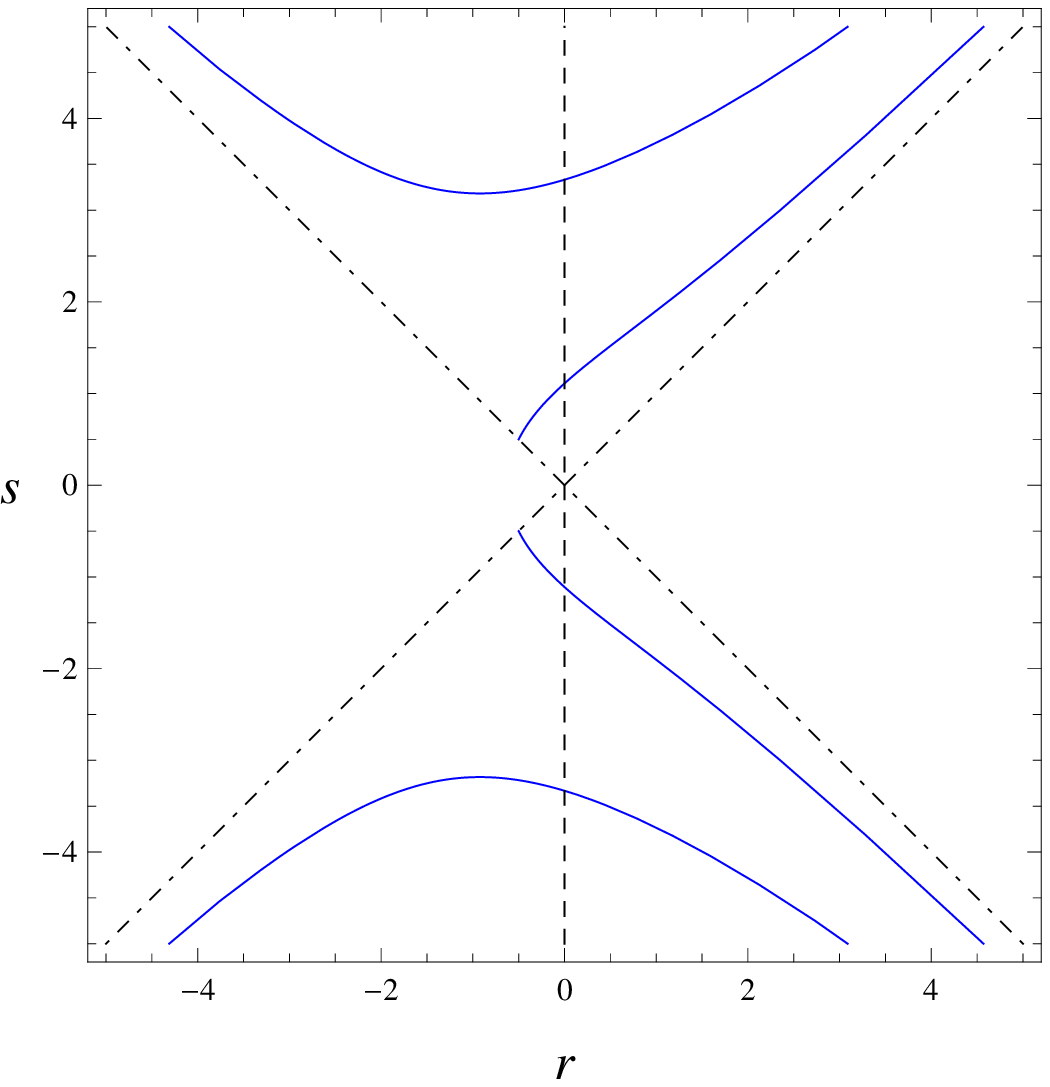}\hspace{1cm}
    \includegraphics[scale=.124,clip]{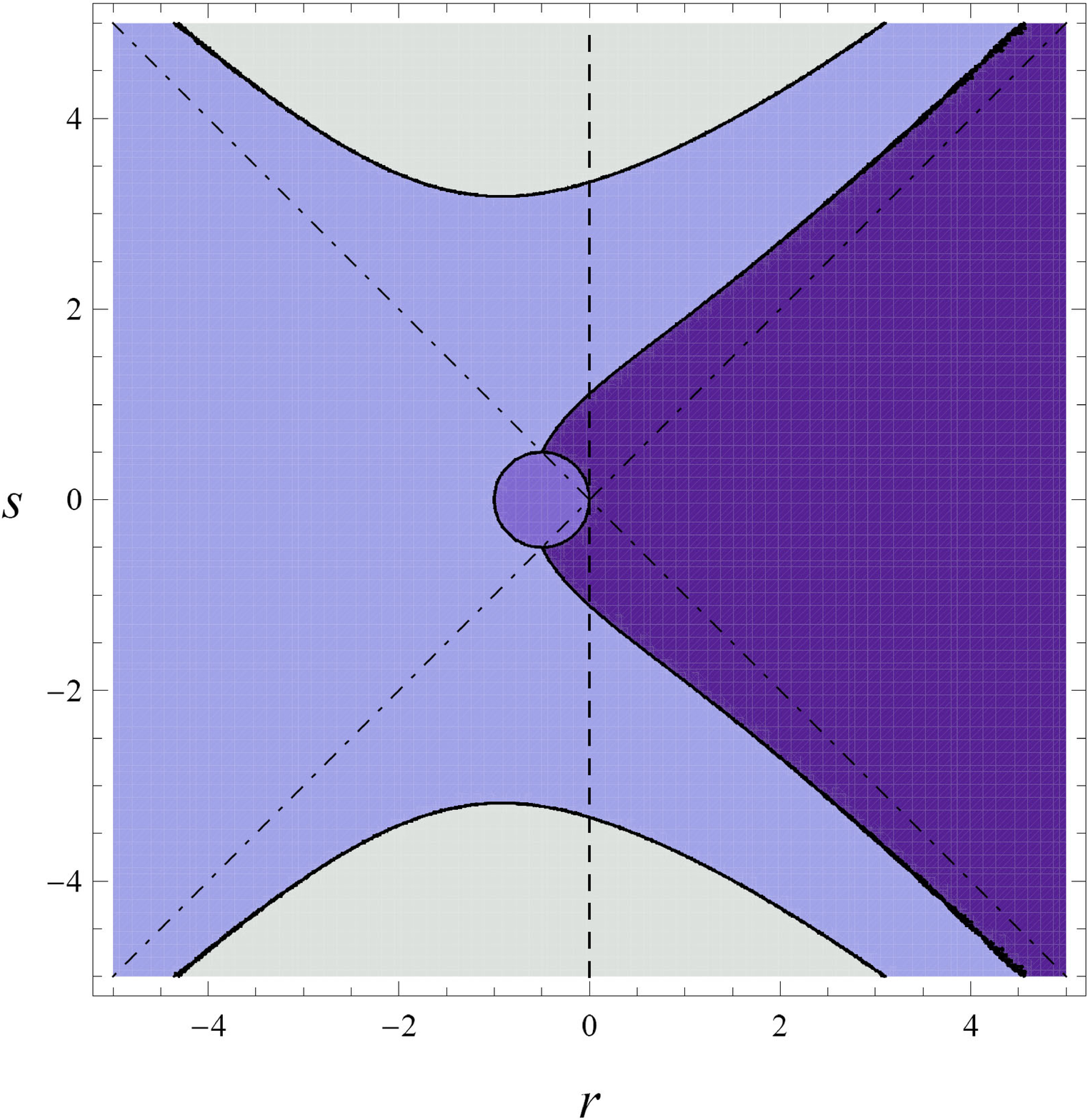}
    \parbox{15cm}{\caption{The curves along which the
    ${\cP\cT}$-symmetric Hamiltonian (\ref{general-h})
    with $\zeta_+=\zeta_-^*=:\zeta$ has spectral singularities
    (the figure on the left) and
    the number of bound states with (the figure on the right).
    As color changes from the darkest to the lightest,
    the number of bound sates takes the values $0, 1, 2, 4$, respectively.
    The vertical dashed lines correspond to the purely imaginary
    coupling $\zeta$, and the diagonal (dotted-dashed) lines are the
    lines $r=\pm s$. There is a bound state within the circle
    $(r+1/2)^2+s^2=1/4$. \label{fig4}}}\end{center}
\end{figure}

Finally, we wish to stress that the above calculations of the metric
operator and the equivalent Hermitian Hamiltonian are reliable only
within the regions in the space of coupling constants that the
Hamiltonian does not posses spectral singularities or bound states
with complex eigenvalues. In Figure~\ref{fig4} we give the location
of the spectral singularities and the number of bound sates for the
general $\cP\cT$-symmetric case.\footnote{Again we have used the
contour integral method described in Ref.~\cite{paper2} to obtain
the location of the spectral methods and the number of bound
states.} In the region painted by the darkest color in the
$r$-$s$-plane (where $r:=\frac{2am\ell}{\hbar^2}\Re(\zeta)$,
$s:=\frac{2am\ell}{\hbar^2}\Im(\zeta)$) the Hamiltonian has no
spectral singularities or bound states. Hence in this region it is
quasi-Hermitian, and (\ref{bounded-metric-1}) gives a reliable
metric operator provided that we stay within the part of this region
that is close to the origin.

\section{Concluding Remarks}

In this article we have employed the pseudo-Hermitian formulation of
quantum mechanics to study a quantum system defined by a Hamiltonian
with two complex point interactions,
$\rH=\rp^2/2m+\zeta_-\delta(\rx+\alpha)+\zeta_+\delta(\rx-\alpha)$.
This requires the construction of an appropriate metric operator
that reveals the structure of the physical Hilbert space and also
the observables of the theory. It further allows for the
construction of an equivalent Hermitian Hamiltonian operator.

The main difficulty one encounters in trying to construct a metric
operator for $H$ is that depending on the choice of the
eigenfunctions of $H^\dagger$, one obtains a ``metric operator''
that may be ill-defined or unbounded. In this article we could
successfully construct a densely-defined and bounded metric operator
to the first order of perturbation for the special cases that the
coupling constants $\zeta_\pm$ differed by a sign. We use this
metric operator to compute the corresponding equivalent Hermitian
operator. This in turn allowed us to compute energy expectation
values for a class of Gaussian wave packets. We then investigated
the physical consequences of the non-Hermiticity of the Hamiltonian
$H$ by examining the contribution of the anti-Hermitian part of the
Hamiltonian (equivalently the nonlocal part of the equivalent
Hermitian Hamiltonian) to the energy expectation values.

In view of the fact that the integral kernel of the metric operator
is a solution of a certain differential equation, we could
generalize our results to the cases that only the imaginary part of
the coupling constants differed by a sign. This allowed for the
application of our results for a general class of double-delta
function potentials that included all $\cP\cT$-symmetric
double-delta function potentials as a subclass. Our investigation of
the physical effects of non-Hermiticity shows that (to the first
nontrivial order of perturbation theory) these effects are not
directly sensitive to the presence of $\cP\cT$-symmetry. This is
because we can easily perturb the real part of the potential in such
a way that the effects of non-Hermiticity is left unaltered while
$\cP\cT$-symmetry is destroyed. Note however that if such a
perturbation does not violate the quasi-Hermiticity of the
Hamiltonian, the Hamiltonian will necessarily possess a symmetry
that similarly to the $\cP\cT$-symmetry is generated by an
invertible antilinear operator \cite{p3,jmp-2003}. This symmetry
cannot however be interpreted as the space-time reflection symmetry.

\section*{Acknowledgments}

This work has been supported by the Scientific and Technological
Research Council of Turkey (T\"UB\.{I}TAK) in the framework of the
project no: 108T009, and by the Turkish Academy of Sciences
(T\"UBA). H. M.-D. has been supported by ``Open Research Center''
Project for Private Universities: matching fund subsidy from MEXT.

\section*{Appendix A: Another Choice for the Wight Functions}

If $\left|\frac{\Im(z_{\pm})}{\Re(z_{\pm})}\right|\ll 1$, and
$\Re(z_{\pm})>0$, the Hamiltonian (\ref{H}) is a quasi-Hermitian
operator \cite{paper2}, and we can use
$\epsilon_{\pm}:=\frac{\Im(z_{\pm})}{\Re(z_{\pm})}$ as perturbation
parameters for a perturbative calculation of a metric operator. This
requires selecting an appropriate set of $|{\phi}_{\fa,
k}^{\vec{z}}\kt$ that would lead to a densely-defined bounded metric
operator. A natural candidate is the following direct generalization
of the expression obtained for a delta function potential with a
complex coupling constant \cite{jpa-2006b}.
    \be
    |{\phi}_{\fa, k}^{\vec{z}}\kt= u(z_+,z_-;k)|\psi_{\fa,
    k}^{\vec{z^*}}\kt,~~~~u(z_+,z_-;k):=\left(1+\frac{z_+z_-}
    {\gamma^2k^2}\right)^{-\frac{1}{2}},
    \label{weight}
    \ee
where $\gamma\in\R^+$ is  arbitrary and
$|\psi_{\fa,k}^{\vec{z^*}}\kt$ are given in (\ref{phi1}) -
(\ref{phi2}).

In view of (\ref{metric}), (\ref{phi2}) and (\ref{weight}),
    \be
    \eta(x,y)=\int_{-\infty}^{\infty}{\psi}_{1, k}^{\vec{z}^*}(x)
    \left({\psi}_{1, k}^{\vec{z}^*}(y)\right)^* W(z_+,z_-,k)
    dk,~~~~W(z_+,z_-,k):=|u(z_+,z_-,k)|^2.
    \label{metric-2}
    \ee
Introducing $\kappa:=\frac{\gamma k }{\sqrt{r_+r_-}}:=\rho k$,
$\varepsilon_1:=\epsilon_++\epsilon_-$, and
$\varepsilon_2:=\epsilon_+\epsilon_-$, we can expand
$W=W(z_+,z_-,k)=W\left(r_+(1+i\epsilon_+),r_-(1+i\epsilon_-),\frac{\kappa
}{\rho}\right)$ in powers of $\epsilon$: \footnote{Here $\epsilon^N$
stands for the sum of terms proportional to
$\epsilon_+^a\epsilon_-^b$ with $a+b=N$.}
    \be
    W=\frac{\kappa^2}{1+\kappa^2}\left\{1-\frac{2\varepsilon_2}{1+\kappa^2}
    +\frac{{\varepsilon}_1^2+{\varepsilon}_2^2}
    {\left(1+\kappa^2\right)^2}\right\}^{-\frac{1}{2}}=
    \frac{\kappa^2}{1+\kappa^2}\left\{1+\frac{\varepsilon_2}{1+\kappa^2}
    -\frac{{\varepsilon}_1^2}{2\left(1+\kappa^2\right)^2}\right\}+\cO(\epsilon^4).
    \label{w-series}
    \ee
Employing this relation in (\ref{metric-2}) and performing the
necessary calculations, we find
    \bea
    \eta(x,y)&=&\eta_{0,0}(x,y)+\eta_{+,+}(x,y)+\eta_{-,-}(x,y)\nn\\&&
    +\eta_{+}(x,y)+\eta_{-}(x,y)+\eta_{-,+}(x,y)+(x\leftrightarrow
    y)^*+\cO(\epsilon^4),
    \label{metric-3}
    \eea
where
    \bea
      &&\eta_{0,0}(x,y)= \frac{1}{2\rho}~\vec{\cE}.
      ~\vec{\cI}^0\left(\frac{x-y}{\rho}\right),\label{eta00}\\
      &&\eta_{\pm,\pm}(x,y)= \frac{\rho }{4}~r_{\pm}^2(1+\epsilon_{\pm}^2)\theta(\pm x^{\mp})
      \theta(\pm y^{\mp}) ~\vec{\cE}.~\left[\vec{\cI}^2\left(\frac{x-y}{\rho}\right)
      -\vec{\cI}^2\left(\frac{x^{\mp}+y^{\mp}}{\rho}\right)\right],\\
      &&\eta_{\pm}(x,y)=\pm \frac{r_{\pm}(1+i\epsilon_{\pm})}{2}\theta(\pm y^{\mp})
      ~\vec{\cE}.~\left[\vec{\cI}^1\left(\frac{x-y}{\rho}\right)
      -\vec{\cI}^1\left(\frac{x^{\mp}+y^{\mp}}{\rho}\right)\right],\\
      &&\eta_{-,+}(x,y)=\pm \frac{\gamma^2}{4 \rho}\theta( y^{-})\theta(-
      x^{+})(1+\varepsilon_2+i(\epsilon_+-\epsilon_-))\times\nn\\
      &&~\vec{\cE}.~\left[\vec{\cI}^2\left(\frac{x^++y^+}{\rho}\right)
      +\vec{\cI}^2\left(\frac{x^-+y^-}{\rho}\right)-\vec{\cI}^2\left(\frac{x-y}{\rho}\right)
      -\vec{\cI}^2\left(\frac{x-y+4a}{\rho}\right)\right],
      \label{eta-m-p}\\
      &&\vec{\cE}:=(1,\varepsilon_2,-\frac{\varepsilon_1^2}{2})=
      (1,\epsilon_+\epsilon_-,-\frac{(\epsilon_++\epsilon_-)^2}{2}),
      ~~~~\vec{\cI}^n:=(\cI_{n,1},\cI_{n,2},\cI_{n,3}), ~~~~n=0,1,2,
      \label{eta-i-j}\\
      && \cI_{n,m}(\alpha):=\frac{1}{2
     \pi}\int_{-\infty}^{\infty}\frac{k^{2-n}e^{ik\alpha}}{(1+k^2)^m}~dk,~~~~~~
     \vec{\cI}_{m}:=(\cI_{0,m}~,~\cI_{1,m}~,~\cI_{2,m}),~~~~m\geq1,~~~~~~\\
     && \vec{\cI}_{1}=\left(\delta(\alpha)
     -\frac{e^{-|\alpha|}}{2}~,~\frac{ie^{-|\alpha|}}{2} {\rm
     sign}(\alpha)~,~\frac{e^{-|\alpha|}}{2}\right),~~~~~~ \vec{\cI}_{2}=\frac{e^{-|\alpha|}}{4}\left(1-|\alpha|~,~i
     \alpha ~,~1+|\alpha|\right),\label{Iij}\\
    && \vec{\cI}_{3}=\frac{e^{-|\alpha|}}{16}\Big(1+|\alpha|-\alpha^2~,~i
     \alpha(1+|\alpha|)~,~3(1+|\alpha|)+\alpha^2\Big)\label{Iij-1},
     \eea
$x^{\pm}$ and $y^{\pm}$ are given by (\ref{x-y-pm}), and
$\left(x\leftrightarrow y\right)$ stands for the sum of the previous
terms with $x$ and $y$ interchanged.

It is not difficult to see that $\vec{\cE}=(1,0,0)+\cO(\epsilon^2)$
and $\vec{\cE}\cdot\vec{\cI}^n=\cI_{n,1}+\cO(\epsilon^2)$. Inserting
the values of $\cI_{n,1}$ given by (\ref{Iij}) - (\ref{Iij-1}) in
(\ref{eta00}) - (\ref{eta-m-p}) and rearranging the terms, we have
    \bea
    \eta(x,y)&=&\frac{1}{2}\delta(x-y)\nn\\
    &&+e^{-\frac{|x-y|}{\rho}}\Big{\{}\frac{-1}{4\rho}
    +\frac{\rho}{8}[r_{+}^2\theta(x^-)\theta(y^-)+r_-^2\theta(-x^+)\theta(-y^+)]
    -\frac{\gamma^2}{8\rho}
    \theta(-x^+)\theta(y^-)(1+i\epsilon_+-i\epsilon_-)\nn\\
    &&\hspace{1.8 cm}+\frac{ir_+}{4} \theta(y^-){\rm sign}(x-y)(1+i\epsilon_+)-
    \frac{ir_-}{4} \theta(-y^+){\rm sign}(x-y)(1+i\epsilon_-)\Big{\}}\nn\\
    &&+{e^{-\frac{|x^-+y^-|}{\rho}}}\Big{\{}-\frac{\rho}{8}r_{+}^2\theta(x^-)\theta(y^-)
    +\frac{\gamma^2}{8\rho}
    \theta(-x^+)\theta(y^-)(1+i\epsilon_+-i\epsilon_-)
    \nn\\
    && \hspace{2.3 cm}
    -\frac{ir_+ }{4}\theta(y^-){\rm sign}(x^-+y^-)(1+i\epsilon_+)\Big{\}}\nn\\
    &&+{e^{-\frac{|x^++y^+|}{\rho}}}\Big{\{}-\frac{\rho}{8}r_{-}^2\theta(-x^+)\theta(-y^+)
    +\frac{\gamma^2}{8\rho}
    \theta(-x^+)\theta(y^-)(1+i\epsilon_+-i\epsilon_-)
    \nn\\
    && \hspace{2.3 cm}
    -\frac{ir_- }{4}\theta(-y^+){\rm sign}(x^++y^+)(1+i\epsilon_-)\Big{\}}\nn\\
    &&-{e^{-\frac{|x-y+4a|}{\rho}}}\Big{\{}\frac{\gamma^2
    }{8\rho}\theta(-x^+)\theta(y^-)(1+i\epsilon_+-i\epsilon_-)\Big{\}}\nn\\
    &&+(x\leftrightarrow y)^*+\cO(\epsilon^2).
    \label{metric-first order}
    \eea
As suggested by this relation, $\eta$ is actually densely-defined
and (perturbatively) bounded, but in the Hermitian limit it does not
tend to the identity operator.

\section*{Appendix B: A Proper Choice for the Wight Functions}
 One can check that
unless one fixes $u_{\pm,\fb}(k)$ in a very special way the metric
operator corresponding to (\ref{phi-till-general}) is an unbounded
operator. In order to obtain such a special choice we consider the
ansatz:
    \bea
    {\phi}_{1, k}^{\vec{z}}(x)&=&{\psi}_{1, k}^{\vec{z^*}}(x)
     + {\phi}_{1, k}^{\fv}(x)+\cO(z^2),
    \label{phi-till-1}\\
    {\phi}_{2, k}^{\vec{z}}(x)&=&{\phi}_{1,-k}^{\vec{z}}(x),
    \label{phi-till-2}
    \eea
where ${\psi}_{\fa, k}^{\vec{z^*}}(x)$ are given by
Eqs.~(\ref{phi1}) - (\ref{phi2}),
     \bea
     {\phi}_{1, k}^{\fv}(x)&:=&\frac{i}{k }\sum_{\lambda=\pm}\sum_{j=1}^{m}
    e^{-ib_j k}z_{\lambda}^*\left(v_{j,1,\lambda}^{*}{\psi}_{1, k}^{\vec{z^*}}(x)+v_{j,2,\lambda}^{*}{\psi}_{2, k}^{\vec{z^*}}(x)\right)
    +\cO(z^2)\nn\\&=&\frac{i}{{\sqrt{2\pi}} }\sum_{\lambda=\pm}\sum_{j=1}^{m}
    \frac{z_{\lambda}^*}{k}\left(v_{j,1,\lambda}^{*}e^{-ik(b_j-x)
    }+v_{j,2,\lambda}^{*}e^{-ik(b_j+x)}\right)+\cO(z^2),
    \label{phi-v-def0}
    \eea
$m$ is a positive integer, and $b_j$ and $v_{j,\fb,\lambda}$ are
respectively real and complex free coefficients. By defining
$\mu(\fb):=2\fb-3$, we can rewrite ${\phi}_{1, k}^{\fv}(x)$ in a
more compact form, namely
     \bea
     {\phi}_{1, k}^{\fv}(x)=\frac{1}{\sqrt{2\pi}}\sum_{\lambda=\pm}\sum_{\fb=1}^{2}\sum_{j=1}^{m}
    \frac{iz_{\lambda}^*v_{j,\fb,\lambda}^{*}}{k }
    e^{-ik[b_j+\mu(\fb)x]}+\cO(z^2).
    \label{phi-v-def}
    \eea

For the cases where (\ref{phi-till-2}) holds, we can use
(\ref{metric}) to express $\eta(x,y)=\br x|\eta|y\kt$ as
    \be
    \eta(x,y)=\int_{-\infty}^{\infty}{\phi}_{1, k}^{\vec{z}}(x)
    \left({\phi}_{1, k}^{\vec{z}}(y)\right)^* dk.
    \label{metric-1}
    \ee
Using
    \be
    \frac{1}{i\pi}\int_{-\infty}^{\infty}\frac{e^{ik\alpha}}{k}\:dk
    = {\rm sign}(\alpha),~~~~\frac{1}{2\pi}
    \int_{-\infty}^{\infty}e^{i k \alpha}\,dk=\delta(\alpha),
    \label{int}
    \ee
and (\ref{phi-till-1}) - (\ref{metric-1}), we then find
    \be
    \eta(x,y)=\delta(x-y)+ \eta^{(1)}(x,y)+\cO(z^2),
    \label{eta-total}
    \ee
where
    \bea
    \eta^{(1)}(x,y)&:=&\eta^{(1)}_+(x,y)+\eta^{(1)}_-(x,y)+
    \eta^{(1)}_{\fv}(x,y)+\left(x\leftrightarrow y\right)^*,
      \label{eta-1}\\
      \eta^{(1)}_+(x,y)&:=&\frac{z_+}{4}
      \left\{ {\rm sign}(x+y-2a)-{\rm sign}(x-y)\right\}\theta(y-a),
      \label{eta-1-p}\\
      \eta^{(1)}_-(x,y) &:=&-\frac{z_-}{4}
      \left\{ {\rm sign}(x+y+2a)-{\rm
      sign}(x-y)\right\}\theta(-y-a),
      \label{eta-1-m}\\
      \eta^{(1)}_{\fv}(x,y)&:=&\sum_{\lambda=
      \pm}\sum_{\fb=1}^{2}\sum_{j=1}^{m}\eta^{(1)}_{j,\fb,\lambda}(x,y)=
      \sum_{\lambda=\pm}\sum_{\fb=1}^{2}
      \sum_{j=1}^{m}\frac{z_{\lambda}
      v_{j,\fb,\lambda}}{2}{\rm sign}(x+\mu(\fb)y+b_j).
      \label{eta-1-mu-j}
       \eea

Note that $\eta^{(1)}_{\fv}(x,y)$ depends on our choice of the free
coefficients $v_{j,\fb,\lambda}$ and $b_j$ in (\ref{phi-v-def}). Our
aim is to make a choice that would lead to a an appropriate metric
operator.

First, we use the identity
    \be
    {\rm sign}(u+v)[{\rm sign}(u)+{\rm sign}(v)]=1+{\rm sign}(u){\rm
    sign}(v),
    \label{sign-peroperty}
    \ee
to obtain
    \bea
    \eta_+^{(1)}(x,y)&=&\frac{z_+}{8}[-2{\rm sign}(x-y)-{\rm
    sign}(x^-+y^-)+1]+\frac{z_+}{4}{\rm sign}(x-y)\theta(-x^--y^-)\nn\\
    &=&\frac{z_+}{8}[{\rm sign}(x^-+y^-)+1]-\frac{z_+}{4}{\rm
    sign}(x-y)\theta(x^-+y^-),
    \label{eta-1-p2}\\
    \eta_-^{(1)}(x,y)&=&\frac{z_-}{8}[2{\rm sign}(x-y)-{\rm
    sign}(x^++y^+)+1]-\frac{z_-}{4}{\rm sign}(x-y)\theta(x^++y^+),
    \label{eta-1-m2}
    \eea
where $x^{\pm}$ and $y^{\pm}$ are given by Eq.  (\ref{x-y-pm}).

Next, we rewrite ${\phi}_{1, k}^{\fv}(x)$ in the form
    \bea
    {\phi}_{1, k}^{\fv}(x)&=&{\phi'}_{1, k}^{\fv}(x)
     -\frac{i}{k\sqrt{2\pi}}\left[\frac{z^*_+}{4}{e^{-ik(x-2a)}}
     -\frac{z^*_-}{4}{e^{-ik(x+2a)}}+\frac{z^*_-}{2}{e^{ikx}}
     \right],\label{phi-v1}
     \eea
where
    \bea
     {\phi'}_{1,k}^{\fv}(x)&:=&\frac{1}{\sqrt{2\pi}}\sum_{\lambda=\pm}\sum_{\fb=1}^{2}\sum_{j=1}^{m'}
    \frac{iz_{\lambda}^*v_{j,\fb,\lambda}^{*}}{k }e^{-ik(b_j+\mu(\fb) x)},\nn
    \eea
and $m'$ is a positive integer. Using the above equations, the first
order term $\eta^{(1)}(x,y)$ of $\eta(x,y)$ can be rewritten as
    \bea
    \eta^{(1)}(x,y)&=&{\eta'}^{(1)}_+(x,y)+{\eta'}^{(1)}_-(x,y)
    +{\eta'}^{(1)}_{\fv}(x,y)+\left(x\leftrightarrow y\right)^*,
    \label{eta-1-2}\\
    {\eta'}^{(1)}_{\pm}(x,y)&=&\frac{z_{\pm}}{8}[1-2{\rm
    sign}(x-y)\theta(x^{\mp}+y^{\mp})],
    \label{eta-1-pm1}\\
    {\eta'}^{(1)}_{\fv}(x,y) &=&\sum_{\lambda=\pm}\sum_{\fb=1}^{2}\sum_{j=1}^{m'}\eta^{(1)}_{j,\fb,\lambda}(x,y)=\sum_{\lambda=\pm}\sum_{\fb=1}^{2}
    \sum_{j=1}^{m'}\frac{z_{\lambda}v_{j,\fb,\lambda}}{2}{\rm sign}(x+\mu(\fb)
      y+b_j)\label{eta-1-v2}\\
    &=&{\eta}^{(1)}_{\fv}(x,y)
    +\frac{z_+}{8}{\rm sign}(x^-+y^-)-\frac{z_-}{8}{\rm sign}(x^++y^+)
    +\frac{z_-}{4}{\rm sign}(x-y).\nn
        \eea
Setting ${\phi'}_{1, k}^{\fv}(x)=0$ or equally
${\eta'}^{(1)}_{\fv}(x,y)=0$ yields
    \be
    {\eta'}^{(1)}(x,y)={\eta'}^{(1)}_{+}(x,y)+{\eta'}^{(1)}_{-}(x,y)+(x\leftrightarrow
    y)^*,
    \label{eta-prim}
    \ee
which is generally not a bounded operator. But if we confine our
attention to the special case where $z_+=-z_-=:z$, we find
    \be
    {\eta'}^{(1)}(x,y)=\frac{i\Im(z)}{2}{\rm
    sign}(x-y)[\theta(x^++y^+)-\theta(x^{-}+y^{-})],
    \ee
which is a bounded operator \cite{ahmet}.

\ed